\documentclass[aps,prx,showpacs,twocolumn,reprint,superscriptaddress]{revtex4-2}

\usepackage{amsmath,amssymb,amsthm,mathrsfs,amsfonts,dsfont,mathtools}
\usepackage{braket}
\usepackage{hyperref}
\usepackage{bm}
\usepackage{enumerate}
\usepackage{siunitx}
\usepackage[table, dvipsnames]{xcolor}
\usepackage{graphicx}
\usepackage{pgf}
\usepackage{pgfplots}
\pgfplotsset{compat=1.16}
\usepackage{tikz}
\usepackage{enumitem}
\usepackage[capitalize]{cleveref}
\usepackage[normalem]{ulem}
\usepackage{physics}
\usepackage[caption=false]{subfig}
\begin{document}
	\title{Space-time tradeoff in networked virtual distillation}
	\author{Tenzan Araki}
	\email{tenzan.araki@physics.ox.ac.uk}
	\affiliation{Department of Physics, Clarendon Laboratory, University of Oxford, Parks Road, Oxford OX1 3PU, United Kingdom}
	\affiliation{Mathematical Institute, University of Oxford, Woodstock Road, Oxford OX2 6GG, United Kingdom}
 \author{Joseph F. Goodwin}
 \affiliation{Department of Physics, Clarendon Laboratory, University of Oxford, Parks Road, Oxford OX1 3PU, United Kingdom}
 \author{B\'alint Koczor}
  \affiliation{Mathematical Institute, University of Oxford, Woodstock Road, Oxford OX2 6GG, United Kingdom}

\begin{abstract}
In contrast to monolithic devices, modular, networked quantum architectures are based on interconnecting smaller quantum hardware nodes using quantum communication links, and offer a promising approach to scalability. Virtual distillation (VD) is a technique that can, under ideal conditions, suppress errors exponentially as the number of quantum state copies increases. However, additional gate operations required for VD introduce further errors, which may limit its practical effectiveness. In this work, we analyse three practical implementations of VD that correspond to edge cases that maximise space-time tradeoffs. Specifically, we consider an implementation that minimises the number of qubits but introduces significantly deeper quantum circuits, and contrast it with implementations that parallelise the preparation of copies using additional qubits, including a constant-depth implementation. We rigorously characterise their circuit depth and gate count requirements, and develop explicit architectures for implementing them in networked quantum systems -- while also detailing implementations in early fault-tolerant quantum architectures. We numerically compare the performance of the three implementations under realistic noise characteristics of networked ion trap systems and conclude the following. Firstly, VD effectively suppresses errors even for very noisy states. Secondly, the constant-depth implementation consistently outperforms the implementation that minimises the number of qubits. Finally, the approach is highly robust to errors in remote entangling operations, with noise in local gates being the main limiting factor to its performance.
\end{abstract}

 \maketitle

\section{Introduction}

\begin{figure*}[ht]
	\centering
	\subfloat[ ]{
		\label{fig: vd_a}
		\includegraphics[width=0.3\textwidth]{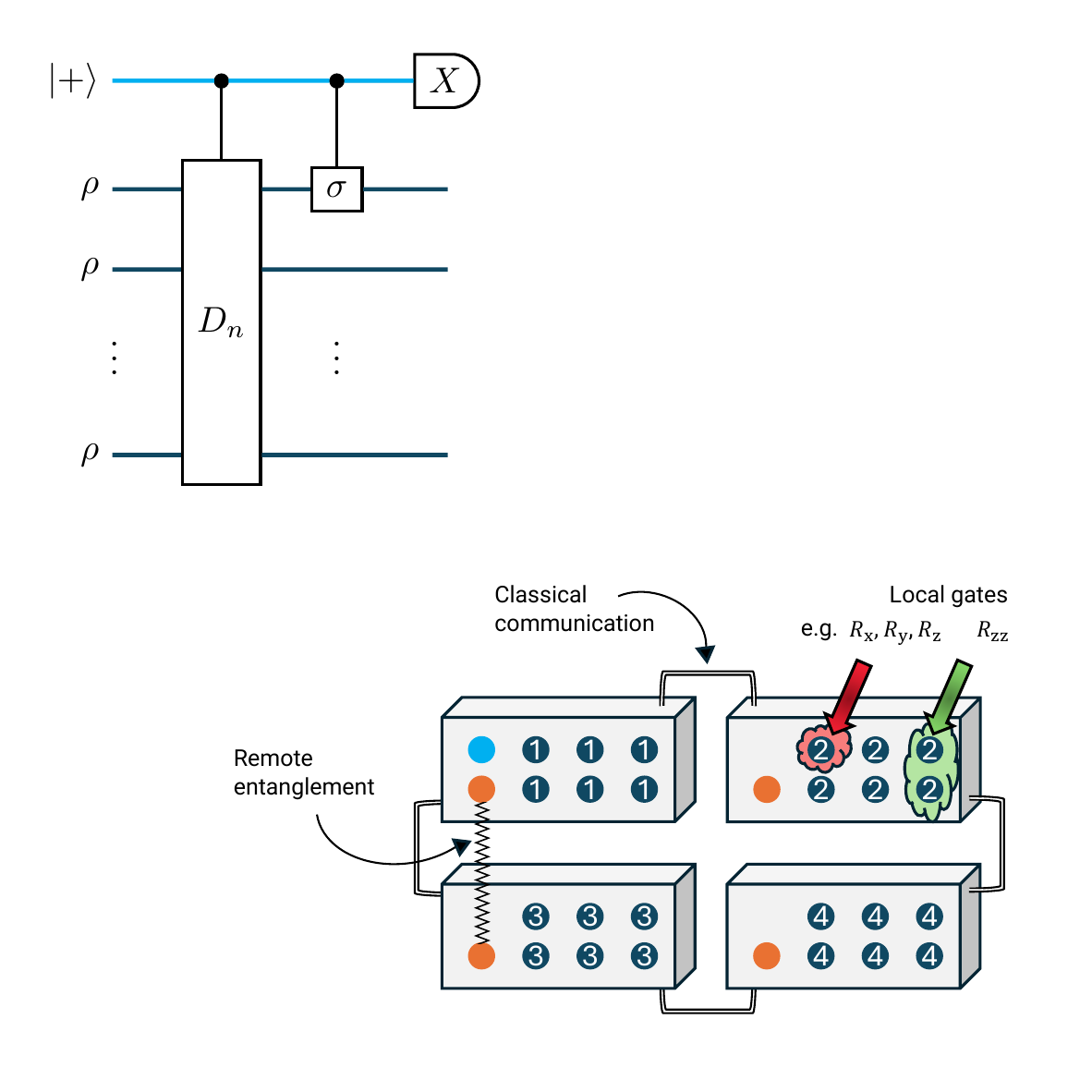}
	}
	\hspace{0.5cm}
	\subfloat[ ]{
		\label{fig: vd_b}
		\includegraphics[width=0.54\textwidth]{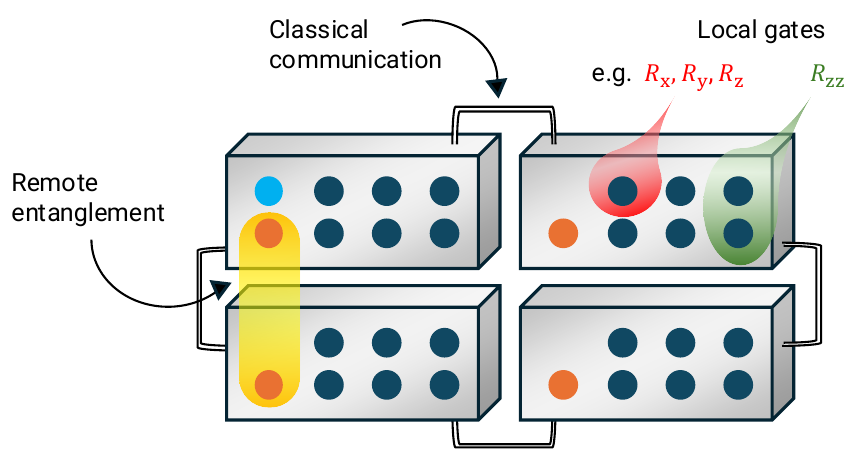}
	}
	\caption{(a) Quantum circuit implementing VD using $n$ copies of the noisy quantum state $\rho$ (dark blue) and a single ancilla qubit (light blue) initialised in $\ket{+}=(1/\sqrt{2})(\ket{0}+\ket{1})$. A derangement operation is applied to all $n$ copies (denoted by $D_n$), conditioned on the state of the ancilla qubit, and is followed by a set of controlled Pauli gates (denoted by $\sigma$) associated with each term in the Pauli basis expansion of $O$. A measurement of the ancilla qubit in the $X$-basis reveals the nonlinear functional Tr$[\sigma\rho^n]$, which is central to the scheme. (b) Conceptual illustration of applying VD with 4 copies of 6 data qubits across a 4-node quantum network, where each copy (dark blue) is placed in a distinct node. The ancilla qubit (light blue) is placed in one of the nodes, and each node has an additional network qubit (orange) used to establish remote entanglement between nodes. With Bell state generation, local operations, and classical communication enabled, VD can be applied across the quantum network.}\label{fig: vd}
\end{figure*}

There has recently been remarkable progress in the development of quantum hardware, but achieving full fault-tolerance remains challenging.
Algorithms that are relatively robust to errors and therefore amenable to near-term quantum hardware have been proposed \cite{cerezo2021variational,bharti2022noisy}, but noise levels on current devices are still limiting their practical efficacy~\cite{zimboras2025myths}.
Solving practical problems beyond system sizes that are classically tractable before achieving full fault-tolerance
requires mitigating errors either in physical gates or logical gates in early fault-tolerant architectures \cite{piveteau2021error,suzuki2022quantum,akahoshi2024partially}. Since one is often concerned with a specific observable property of a quantum system in typical applications of quantum computers, a broad range of quantum error mitigation (QEM) \cite{cai2023quantum} techniques have been proposed that suppress errors in observable expectation values by postprocessing results from an ensemble of noisy quantum circuits -- as opposed to fully correcting all errors in the quantum states.

Suppose we want to obtain the expectation value of an observable $O$ with respect to some quantum state $\ket{\psi_{\text{id}}}$ (i.e., $\bra{\psi_{\mathrm{id}}}O\ket{\psi_{\mathrm{id}}}$ or equivalently Tr$[O\ketbra{\psi_{\mathrm{id}}}]$) that would be prepared by an ideal noise-free quantum circuit. In practice, due to noise in the quantum hardware, the output of the circuit is a mixed state $\rho$, 
and an expected value measurement therefore yields Tr$[O\rho] \neq \bra{\psi_{\mathrm{id}}}O\ket{\psi_{\mathrm{id}}}$. Virtual distillation (VD) \cite{huggins2021virtual, koczor2021exponential} applies a shallow entangling operation (known as a controlled ``derangement'') to $n$ copies of the noisy state $\rho$ as illustrated in \Cref{fig: vd_a}, where each copy is prepared independently by applying the same noisy state preparation circuit \footnote{In practice, even if identical gates are applied, the copies may not be identically prepared due to variations in the noise channels. VD remains effective as long as $\ket{\psi}$ is the dominant eigenvector across all copies, and its performance is limited by the lowest dominant eigenvalue amongst the copies as shown in \cite{koczor2021exponential}.}. Then, measuring an ancilla qubit gives rise to an estimator that approaches $\langle \psi | O | \psi \rangle$ exponentially with respect to $n$. Here, $\ket{\psi}$ is the dominant eigenvector of $\rho$, which is subject to a slight deviation from the ideal state, i.e., $|\bra{\psi} \ket{\psi_{\mathrm{id}}} | \neq 1 $, known as a coherent mismatch~\cite{koczor2021the}. In typical practical scenarios, this is negligible compared to the incoherent loss of fidelity -- where the latter is suppressed exponentially effectively by VD.

While coherent errors are not suppressed by VD, they can be addressed either by Pauli twirling \cite{silva2008scalable, magesan2012characterizing, cai2019constructing, cai2020mitigating} or by combining it with other QEM techniques \cite{yoshioka2022generalized,yang2024dual,koczor2024sparse,koczor2024probabilistic}. A more serious drawback of VD is the considerable overhead in the total number of qubits due to the preparation of multiple copies, a significant concern given that scaling qubit number remains the foremost challenge for almost all proposed quantum computing platforms. The problem can be alleviated by combining VD with dual-state purification \cite{huo2022dual, cai2021resource}, in which case the number of qubits can be halved. A qubit-efficient implementation of VD was also proposed \cite{czarnik2021qubit}, such that only two quantum registers are used to store $n$ copies by repeatedly preparing new copies and resetting those that are no longer needed. However, while reducing the number of qubits required, such implementations lead to deep circuits which suffer from increased idling errors. In the present work, we rigorously analyse circuit depth and gate count requirements of a variety of such implementations, including a variant that achieves a constant circuit depth, albeit requiring an increased number of qubits.

With VD protocols requiring at least double the number of qubits versus unmitigated circuits, consideration of practical routes to scaling and their corresponding impact on the implementation of the VD routine are essential in predicting real-world performance. In widely-used trapped-ion quantum architectures, e.g., scaling via producing a long chain of ions results in slower entangling gate speeds and lower fidelities, due to increasingly crowded motional mode frequencies; meanwhile, instead using many spatially separated small chains increases ion shuttling overheads and control hardware complexity.
Beyond the barriers to scaling qubit number, monolithic quantum architectures pose challenges for implementing VD and other large-scale protocols due to, e.g, connectivity constraints. Therefore, in this work we consider applying VD across a modular network of quantum processing units (QPUs) \cite{gottesman1999demonstrating, monroe2014large, jiang2007distributed,caleffi2024distributed} as illustrated in \Cref{fig: vd_b}.

Coherent quantum operations across distinct nodes currently represent a bottleneck in modular quantum architectures due to their relatively slow and noisy operations. However, typical practical applications, e.g., simulating time-evolution via Trotterisation, require state preparation circuits that are significantly---potentially orders of magnitude---deeper than the remote entangling operations required for VD which are applied across the entire network
-- VD is therefore well-suited to modular architectures~\cite{jnane2022multicore}.
While in this work we illustrate applications of VD in noisy devices, we will detail that our implementations
are immediately compatible with early fault-tolerant quantum devices, whereby applying QEM 
techniques for mitigating residual, non-negligible logical errors is expected to be a major enabler
-- and VD is expected to be one of the most promising candidates for the early fault-tolerant regime~\cite{zimboras2025myths}.

In \Cref{sec: practical}, we describe three general implementations of VD that represent edge cases in achieving space-time tradeoffs. We then introduce quantum networks and detail architectures for implementing VD in \Cref{sec: applying}. We also briefly discuss implementation details of VD in modular early fault-tolerant quantum architectures in this section. In \Cref{sec: numerical}, we demonstrate simulation results for comparing the performance of different VD implementations in a prototypical example of a quantum network of trapped-ion QPUs, considering errors that accumulate during initialisation, idling time, detection, and local and remote gate operations. We finally conclude in \Cref{sec: conclusion}.

\section{Space-time tradeoffs in VD implementations \label{sec: practical}}
Given $n$ copies of a quantum state $\rho$ and some Pauli string $\sigma$, VD can be performed
via a Hadamard test as illustrated in \Cref{fig: vd_a}, where sampling the ancilla qubit allows one to estimate Tr$[\sigma\rho^n]$.
Doing so for the individual terms $\sigma_i$ in the Pauli decomposition of an observable $O = \sum_i c_i \sigma_i$ 
leads to an approximation of the ideal expectation value as
\begin{equation}
     \sum_i c_i \frac{\text{Tr}[\sigma_i\rho^n]}{\text{Tr}[\rho^n]} = \langle \psi | O | \psi \rangle + \mathcal{E},
    \label{eq: vd}
\end{equation}
where $\mathcal{E}$ is an error that decays exponentially with respect to $n$ and $\ket{\psi}$ is the dominant eigenvector of $\rho$. Please refer to refs~\cite{huggins2021virtual} and~\cite{koczor2021exponential} for proofs. In typical practical scenarios, $\ket{\psi}$ is a good approximation of the ideal quantum state~\cite{koczor2021the}, such that
$\langle \psi_{\text{id}} | O | \psi_{\text{id}} \rangle \approx \langle \psi | O | \psi \rangle$.

\begin{figure}[tb]
    \centering
    \includegraphics[width=0.48\textwidth]{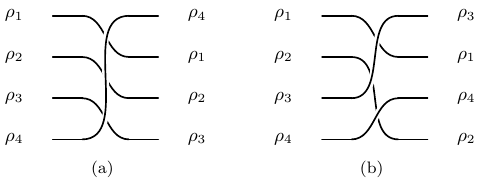}
    \caption{$n$-cycles applied as (a) cyclic rotation and (b) brickwork, shown for the example where $n = 4$.}\label{fig: derange}
\end{figure}

\begin{table*}[tb]
	\centering 
	\begin{tabular}{@{\hspace{2mm}}l@{\hspace{7mm}}c@{\hspace{9mm}}c@{\hspace{10mm}}c@{\hspace{2mm}} }
		\\[-3mm]
		\hline\hline
		\\[-4.5mm]
		& \textbf{QECR} & \textbf{CR} & \textbf{BW} \\ 
		\hline
		\,\textbf{Register count} & $2$ & $n$ & $n$ \\ 
		\,\textbf{Qubit count} & $2N+1$ & $nN+1$ & $nN + \lfloor\frac{n-1}{2}\rfloor$ \\
		\,\textbf{C-SWAP count} & $(n-1)N$ & $(n-1)N$ & $\lfloor\frac{n-1}{2}\rfloor N$ \\
		\,\textbf{BSM count} & $N$ & $N$ & $\lfloor\frac{n}{2}\rfloor N$ \\
		\,\textbf{Circuit depth} & $(n-1)d_{\rho}+(n-2)d_{\text{S}}+d_{\text{B}}$ & $d_{\rho}+(n-2)d_{\text{S}}+d_{\text{B}}+d_{\sigma}$ & $d_{\rho}+d_{\text{S}}+d_{\text{B}}+d_{\sigma}$
		\\[2mm] \hline \hline
	\end{tabular} 
    \caption{Quantum resources required for QECR, CR, and BW for general architectures, showing a space-time tradeoff. C-SWAP and BSM counts are given by the number of individual operations rather than the number of transversal set of operations. The bottom row shows the total circuit depth in terms of the circuit depth of preparing each copy ($d_{\rho}$) and applying transversal C-SWAP gates ($d_{\text{S}}$), transversal BSMs ($d_{\text{B}}$), and the controlled Pauli gates ($d_{\sigma}$). We assume that $d_{\rho}$ is much greater than $d_{\sigma}$ and the circuit depth required to prepare a GHZ state.}\label{tab:tradeoff_table}
\end{table*}

The derangement operation, denoted as $D_n$ in \Cref{fig: vd_a}, is an $n$-cycle permutation of the $n$ copies of $\rho$, which can be decomposed non-uniquely into a minimum of $n{-}1$ pairwise swaps \cite{denes1959representation}.
Therefore, the circuit requires $n{-}1$ controlled SWAP (C-SWAP) operations applied transversally across each pair of $N$-qubit registers,
and there is significant flexibility in choosing a specific permutation pattern. C-SWAP gates can also be implemented in parallel by introducing additional ancilla qubits, which was studied in ref.~\cite{liang2023unified} as a space-time tradeoff with respect to both $N$ and $n$. In this work, however, we restrict our attention to the space-time tradeoff in $n$ by applying a sequential implementation controlled by a single ancilla qubit, as this is the most relevant case to the networked architecture that we consider (\Cref{sec: applying}).

In the following, we present three practical implementations of VD considering general hardware architectures (i.e., either monolithic or modular), where the minimal number of $n{-}1$ transversal C-SWAP gates are applied. The implementations are each associated with one of the two derangements shown in \Cref{fig: derange}, which are called \textit{cyclic rotation} and \textit{brickwork} in this work. We also note that the number of distinct $n$-cycles grows rapidly with the number of copies as $(n-1)!$ \cite{koczor2021exponential}, but here we focus on three particular choices to explore edge cases in space-time tradeoffs, summarised in \Cref{tab:tradeoff_table}. We anticipate that the space-time tradeoff in the intermediate regime varies monotonically between these extremes.

In order to minimise circuit depth, we apply each implementation destructively by replacing the final layer of transversal C-SWAP gates with a layer of transversal Bell State Measurments (BSMs), i.e., pairs of qubits are measured in the basis spanned by Bell states, as detailed in \Cref{sec: rep}. Since C-SWAP gates can rarely be implemented natively in existing hardware platforms, trading them with BSMs can lead to significantly shallower circuits after compilation. As we will show, this trick is particularly effective when applied to the below BW, where half of the C-SWAP gates are in the final layer, and can therefore be removed, along with a slight yet experimentally crucial reduction in the size of the multipartite entangled state that the implementation utilises. While further, potentially radical reduction in quantum resources is possible by exploiting the locality of entanglement in certain many-body systems \cite{hakoshima2024localized}, we do not make use of those schemes as our aim is to assess the performance of VD in a general setting.

\begin{figure*}
    \centering
    \includegraphics[width=0.8\textwidth]{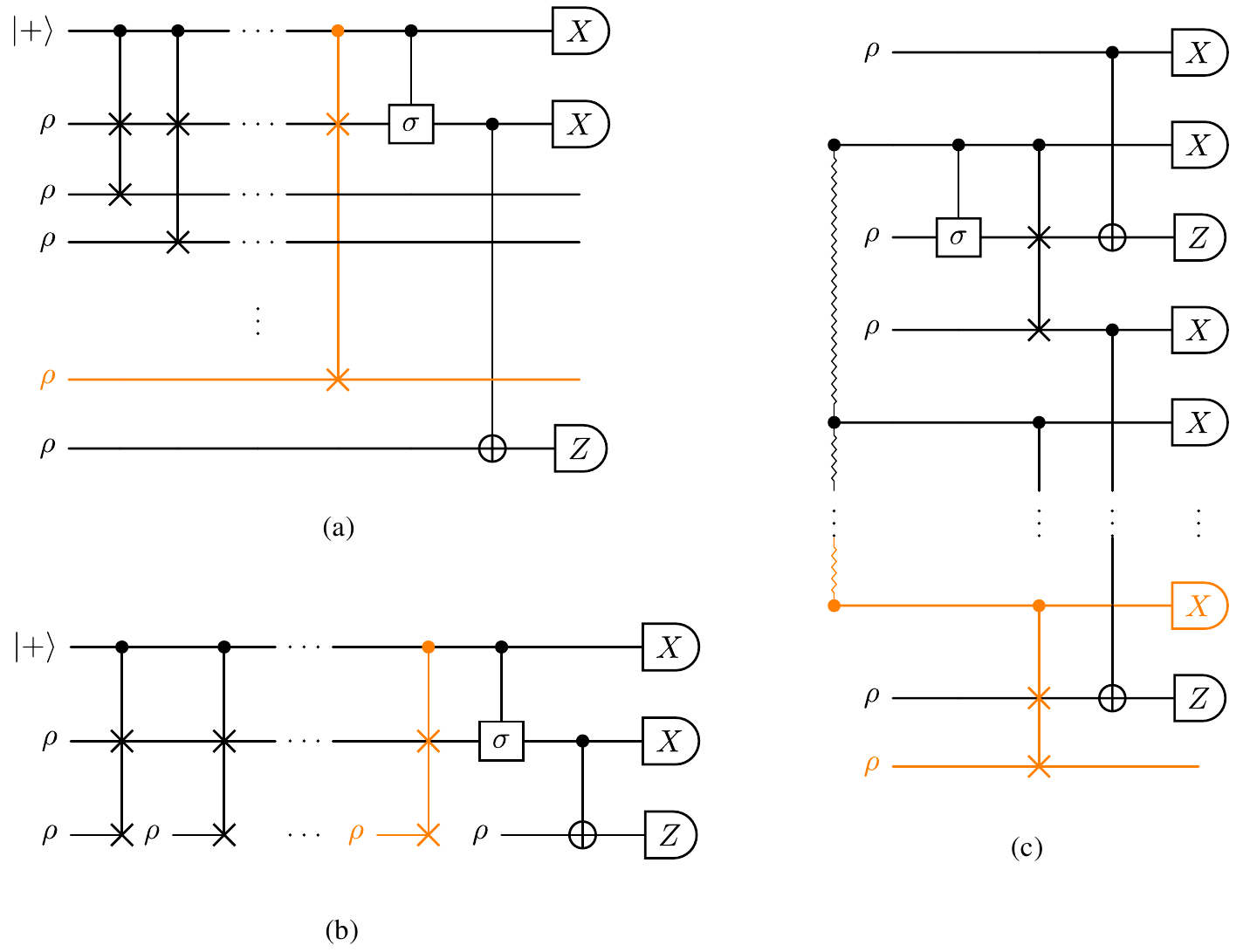}
    \caption{Three implementations of VD after replacing the final layer of C-SWAP gates with BSMs: (a) Cyclic rotation (CR), (b) qubit-efficient cyclic rotation (QECR), and (c) brickwork (BW). The components highlighted in orange are present (absent) when the number of copies is odd (even). In every case, the controlled derangement in \Cref{fig: vd_a} is decomposed into C-SWAP gates and BSMs, where the latter consists of a CNOT gate followed by $X$- and $Z$-basis measurements. The wavy line across multiple ancilla qubits represents a GHZ state. The amount of quantum resources used in each implementation is summarised in \Cref{tab:tradeoff_table}.}
    \label{fig: vd_circs}
\end{figure*}

\noindent\textbf{Cyclic rotation (CR):} The most commonly applied derangement for VD is shown in \Cref{fig: derange}a, where a set of noisy copies initially prepared in the order $\{\rho_1, \rho_2, \rho_3, ..., \rho_n\}$ undergoes a permutation to $\{\rho_n, \rho_1, \rho_2, ..., \rho_{n-1}\}$. While various arrangements of pairwise swaps can give rise to such cyclic rotation, here we consider sequentially swapping between registers $1$ and $k$, where $k$ increments from $2$ to $n$. Due to the sequential nature, only a single set of transversal C-SWAP gates can be applied destructively: the final set. The quantum circuit with its final C-SWAP gates applied destructively is shown in \Cref{fig: vd_circs}a. CR requires in total $nN+1$ qubits, i.e., $n-1$ copies of an $N$-qubit register and a register with $N+1$ qubits to account for the single ancilla qubit. This implementation then requires $(n-1)N$ C-SWAP gates and $N$ BSMs.\\

\noindent\textbf{Qubit-efficient cyclic rotation (QECR):} By noticing that most registers do not participate in subsequent operations after being swapped in CR, a qubit-efficient version of CR has also been proposed \cite{czarnik2021qubit}, where the same derangement can be applied with a qubit number that is constant in the number of copies, as shown in \Cref{fig: vd_circs}b. While CR prepares the copies in parallel, the latter does so by resetting the data register that is no longer in use and re-preparing new copies in it. Noting that the preparation of new copies is typically the deepest part of the quantum circuit, the drawback of QECR is the additional circuit depth and the resulting accumulation of idling errors. Only $2N+1$ qubits are required in QECR, consisting of $N$ and $(N+1)$-qubit registers. Due to the same derangement applied, the number of C-SWAP gates and BSMs is the same as in CR.\\

\noindent\textbf{Brickwork (BW):} While both implementations described above lead to circuit depths that scale linearly in $n$, VD can also be applied in a constant circuit depth \cite{liang2023unified,quek2024multivariate} by considering the derangement shown in \Cref{fig: derange}b. In this third implementation, $2$ layers of transversal C-SWAP gates are applied in a brickwork structure for any $n$. The resulting quantum circuit, shown in \Cref{fig: vd_circs}c, involves $\lfloor({n-1})/{2}\rfloor$ ancilla qubits that share a Greenberger–Horne–Zeilinger (GHZ) state in order to parallelise the transversal C-SWAP gates. In \Cref{sec: gat}, we show that the circuit construction can be viewed as fanning out \cite{haner2021distributed} the ancilla qubit and fanning it back in. While also benefiting from the parallelism in the preparation of new copies seen in CR, the BW implementation could furthermore lead to much shallower circuit depths in practice. This is not only due to the constant-depth nature of the construction, but also because half of the transversal C-SWAP gates (specifically, all of those in the second layer) can be replaced by BSMs. As a result, this implementation involves a total of $nN+\lfloor({n-1})/{2}\rfloor$ qubits, $\lfloor({n-1})/{2}\rfloor N$ C-SWAP gates, and $\lfloor{n}/{2}\rfloor N$ BSMs. Each qubit of the GHZ state is located in every other register, such that only $N$ or $N+1$ qubits is required in each of the $n$ registers.

\section{Distributed implementations}\label{sec: applying}

In distributed, networked quantum computing architectures, multiple QPUs of moderate scale are interconnected by quantum (and classical) communication links. Each QPU, equipped with the usual initialisation, local gate, and measurement operations, is a \textit{module} or \textit{node} of the quantum network. Crucially, quantum gates can be applied remotely across pairs of nodes by consuming Bell states prepared between them \cite{main2025}, enabling universal quantum computation across the quantum network. One or more qubits per node are typically dedicated to storing a half of a Bell pair; these are called \textit{network qubits}. In this work, we refer to the qubits on which the main computation is performed (i.e., those storing the copies $\rho$ in VD) as \textit{data qubits}. The controlled derangement operation is applied on these data qubits, conditioned on the \textit{ancilla qubit}.

Although promising experimental progress has been made in the remote preparation of Bell states \cite{stephenson2020high,krutyanskiy2023,saha2025}, they are and will likely remain a bottleneck for such architectures with regards to achievable rate and fidelity versus local bipartite entangling gates. In this view, the structures of CR, QECR, and BW implementations of VD are well-suited for implementation in a distributed architecture: by placing each copy in a different node, the deepest part of the computation (the preparation of copies) can be performed independently within each node, and only a small number of remote operations bridge them together in the end. Therefore, in this work, we study the performance of VD across a quantum network when placing one copy per node at any given time and at most one ancilla qubit per node.

\subsection{Connectivity\label{subsec: connectivity}}
\begin{figure}
    \centering
    \includegraphics[width=0.45\textwidth]{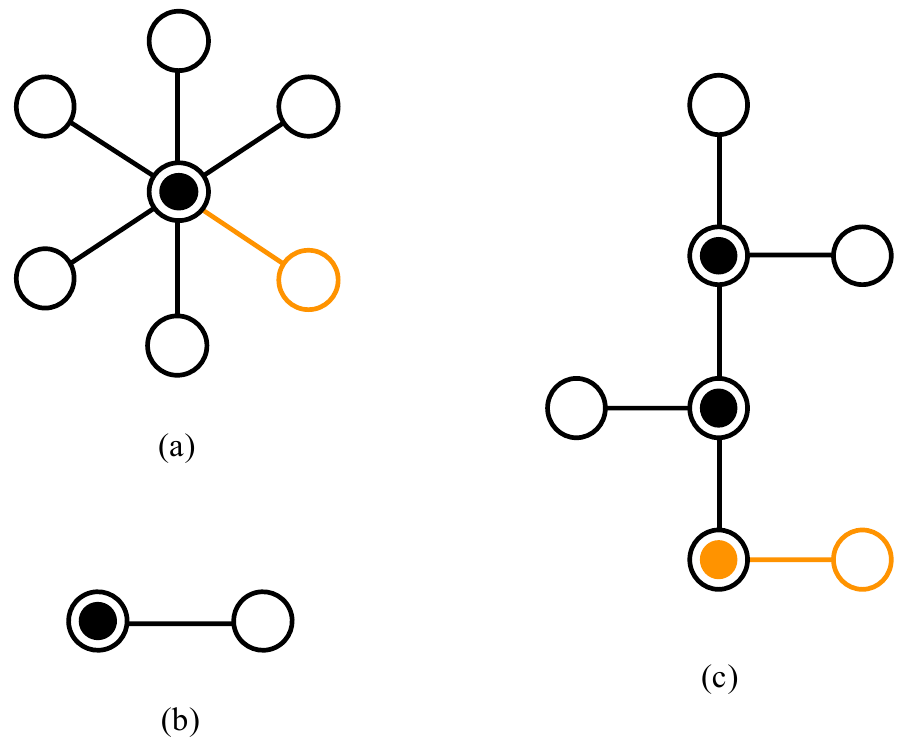}
    \caption{Connectivities between nodes required to apply VD with 6 or 7 copies across a quantum network, where one copy is stored in each node at every moment, using (a) CR (b) QECR and (c) BW implementations. Each white circle represents a node containing a single copy, and those with a black solid circle also includes an ancilla qubit. Bell states can be prepared between each pair of nodes that are connected by a line. Components highlighted in orange are present (absent) when considering 7 (6) copies.}
    \label{fig: vd_connectivity}
\end{figure}

The connectivity requirement of the quantum network depends on the implementation considered, and centers around the nodes that contain the ancilla qubits. For clarity, we refer to these nodes that contain ancilla qubits as \textit{control nodes} and the others as \textit{target nodes}. If CR is applied across a quantum network, remote operations will be applied between a single control node and all the target nodes, leading to a star graph connectivity (\Cref{fig: vd_connectivity}a). If QECR is chosen instead, only 2 nodes are sufficient for any number of copies, consisting of one control node and one target node (\Cref{fig: vd_connectivity}b). For BW, the $\lfloor ({n-1})/{2} \rfloor$ ancilla qubits that collectively share a GHZ state are all placed in different control nodes, requiring at least a linear connectivity between them. If $n$ is odd, one of the two outermost control nodes is connected to two target nodes, and each of the other control nodes is connected to only one target node. It is similar in the case where $n$ is even, but the outermost control node on the other end would also be connected to two target nodes (\Cref{fig: vd_connectivity}c).

\subsection{Remote operations\label{subsec: remote}}
\begin{figure*}[ht]
    \centering
    \includegraphics[width=0.9\textwidth]{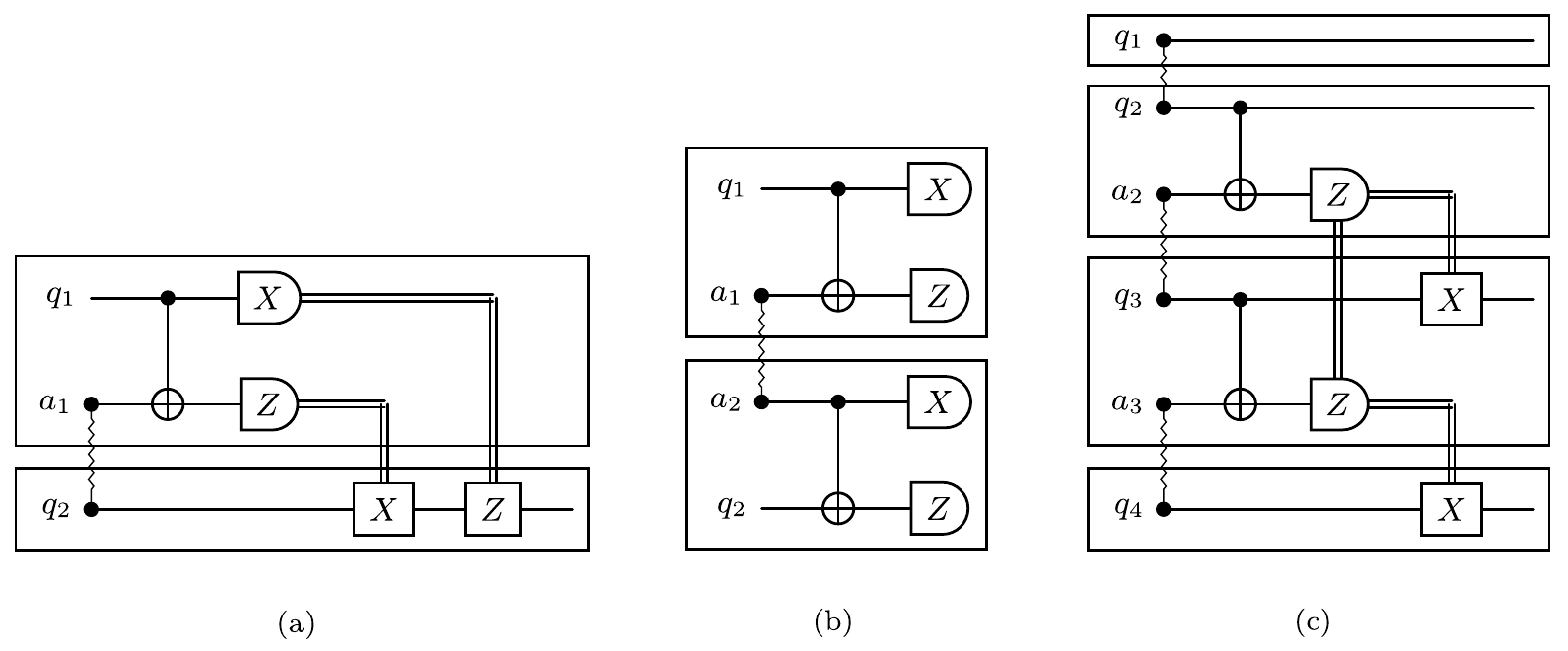}
    \caption{Remote operations used to apply VD across a quantum network. (a) Teleportation of the state of qubit $q_1$ to a qubit $q_2$ in a different node, given a Bell pair shared between qubits $a_1$ and $q_2$. (b) BSM between qubits $q_1$ and $q_2$ in different nodes. A remote BSM outcome of both 1 corresponds to when both pair of qubits $(q_1,a_2)$ and $(a_1,q_2)$ produce odd parity outcomes. (c) Preparation of an $n$-qubit GHZ state in constant time -- example for $n=4$ where the qubits $q_1$, $q_2$, $q_3$, and $q_4$ are located in distinct nodes and ancilla qubits $a_2$ and $a_3$ are used. The $X$ gates are corrections applied based on the collective outcome of the $Z$-basis measurements. The scheme scales straightforwardly to $n>4$ qubit GHZ states.}\label{fig: remote operations}
\end{figure*}

The following three remote operations, illustrated in \Cref{fig: remote operations}, are required when applying VD across a quantum network using the implementations described in \Cref{sec: practical}.\\

\noindent \textbf{Remote C-SWAP gate:} One of the copies to be swapped is placed in a control node and the other in a target node. While it is possible to decompose the C-SWAP gate into a series of single- and two-qubit gates and apply the two-qubit gates remotely, such an attempt would require a suboptimal number of Bell state preparations that in general cannot be parallelised. A simpler method considered in this work is to use quantum teleportation, which transfers the state of the qubit in the target node to the control node (see \Cref{fig: remote operations}a), such that the C-SWAP gate can be applied locally within the control node. Note that in all implementations, we only teleport a copy to the control node but never teleport it back. For CR and QECR, this is because the copy that undergoes teleportation does not participate in the rest of the computation after the C-SWAP gate. While most of the copies involved in the C-SWAP layer of BW are still needed in the operations that follow, the subsequent operations can be performed from within the control node. Thus, each remote C-SWAP gate can be applied by consuming a single Bell pair prepared across one network qubit in each of the two nodes.\\

\noindent \textbf{Remote BSM:} This is applied in the second layer of BW. We consider the implementation shown in \Cref{fig: remote operations}b, which corresponds to applying local BSMs after sharing a single Bell pair between the two participating nodes. The qubits on which remote BSM is applied do not need to be teleported to a different node after the C-SWAP layer of BW.\\

\noindent \textbf{Remote GHZ state preparation:} This is required to achieve the highly parallel structure of BW, and can be realised in various ways (see, e.g., \cite{debone2020protocols, ainley2024multipartite}). In particular, a GHZ state of $N_{\text{GHZ}}$ qubits can be prepared in a constant circuit depth with the help of at most one network qubit per node using $N_{\text{GHZ}}-1$ Bell states, along with local gates and adaptive operations, as shown in \Cref{fig: remote operations}c. The network qubits used in this step can be erased and reused to apply VD, such that no additional network qubits are required.

\subsection{Fault-tolerant architectures\label{subsec: error}}
While our numerical simulations will assume implementations in physical qubits using physical gates, all space-time tradeoff considerations apply equally to logical qubits and logical operations. It has recently been argued that VD is one of the most promising candidates for error mitigation in the early fault-tolerant era, whereby logical qubits are in limited supply and have non-negligible logical error rates~\cite{zimboras2025myths}. Applying VD to mitigate the effect of residual logical errors in expected value measurements may allow---in certain error regimes---for a more efficient noise reduction than committing the increased number of physical qubits to increase the code distance. Furthermore, exploiting modular quantum architectures for VD remains equally relevant in the early fault-tolerant context, such that VD can be combined with error-corrected qubits similarly as concatenating QEC codes in distributed schemes.

When QEC is applied across a quantum network, the error correction cycle will be enacted remotely across multiple nodes, each holding data qubits to be protected. These data qubits could be individual physical qubits, but may even be logical qubits that are concatenated by the code on the higher layer. Syndromes can be extracted using single ancillary registers in the standard approach, but a more architecture-native method involves measuring parities through GHZ states \cite{nickerson2013topological,nickerson2014freely}. Entanglement distillation \cite{bennett1996purification,bennett1996mixed,dur2003entanglement,jiang2007distributed,yamamoto2024virtual} can be applied either physically or logically to improve the fidelity of remotely established Bell pairs.

Consider the same distributed implementation of VD as discussed in the rest of this section, but applied to logical qubits instead of physical ones. Then, each copy $\rho$ is encoded logically and may deviate from $\ket{\psi_{\text{id}}}$ due to, e.g.,
residual logical errors or compilation errors in randomised compiling \cite{campbell2019random,kiumi2024te,kliuchnikov2023shorter,koczor2024sparse}.
The controlled derangement operation can be implemented fault-tolerantly and remotely across the quantum network with the use of network qubits -- the required transversal C-SWAP gates can similarly be implemented by teleporting single logical qubits from the target node to the control node, and by applying local logical C-SWAP gates, each of which can be  decomposed into a sequence of a small number of Clifford+T 
gates~\cite{selinger2013quantum,amy2013meet}.

\section{Numerical simulations \label{sec: numerical}}

In the following, we numerically compare and analyse the performance of the three practical implementations of VD described in \Cref{sec: practical} when applied across a quantum network. In particular, we consider a noisy quantum network of trapped-ion QPUs as a concrete example. Moreover, we consider a practically motivated benchmark problem whereby we estimate an observable expectation value after time-evolving a simple initial state under a random-field Heisenberg model \cite{luitz2015many,childs2018towards} using Trotterisation \cite{trotter1959product,suzuki1976generalized} (see \Cref{subsec: ran} for details). Thus, the simulations involve the practical implementations shown in \Cref{fig: vd_circs} compiled to the trapped-ion quantum network, with input states $\rho$ corresponding to noisy realisations of Trotter circuits.

\subsection{Details of noise models \label{subsec: simulation}}

\begin{figure*}
	\centering
	\includegraphics[width=\textwidth]{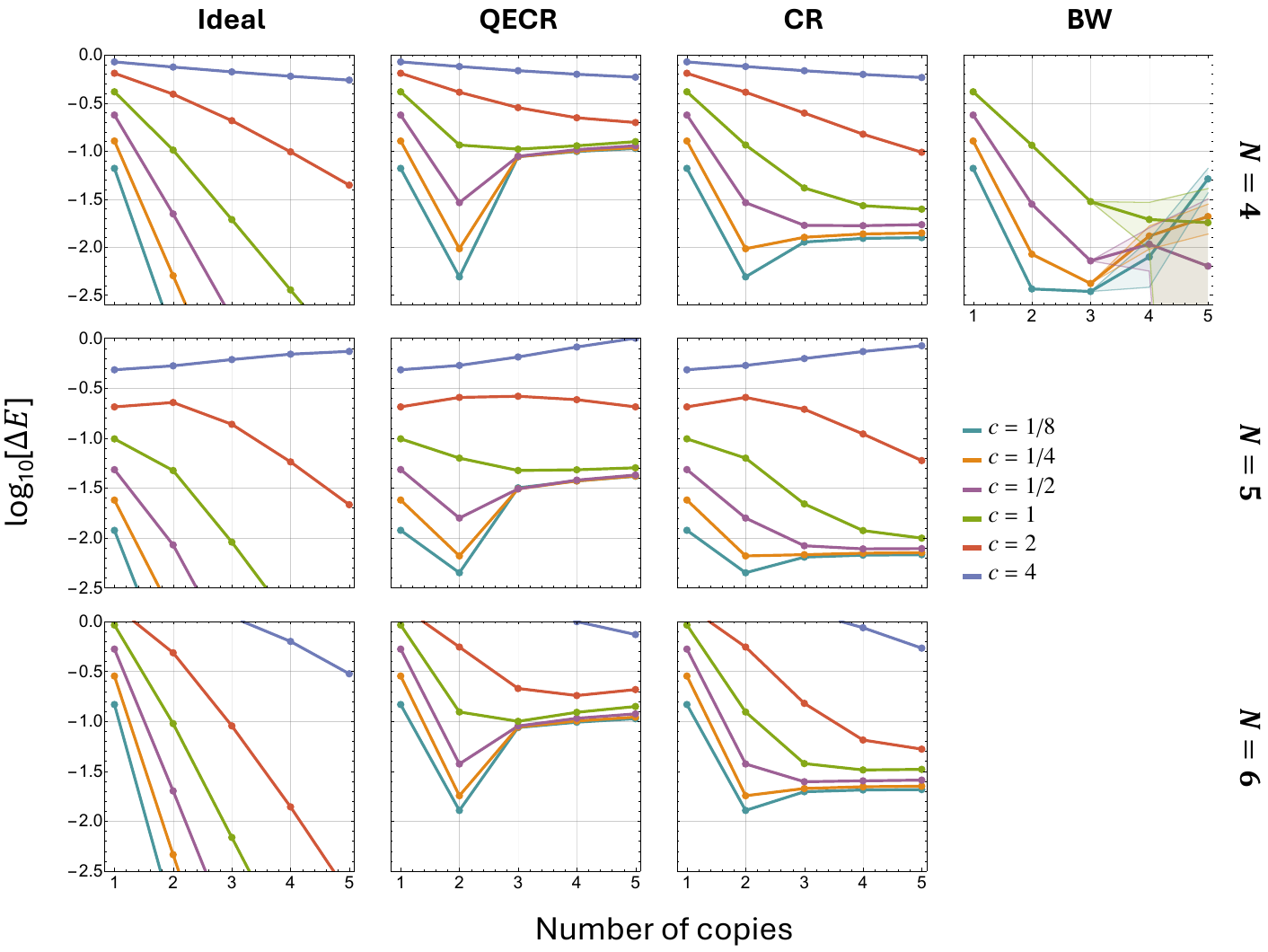}
	\caption{Comparison between an ideal VD and QECR, CR, and BW (columns) for system sizes $N \in \{4,5,6\}$ (rows) and error probabilities $\{p_{\text{1Q}}, p_{\text{2Q}}, p_{\text{Bell}}\} = \{c p^*_{\text{1Q}}, c p^*_{\text{2Q}},c p^*_{\text{Bell}}\}$ controlled by the parameter $c \in \{1/8, 1/4, 1/2, 1, 2, 4\}$ (line colors). In each case, the base-10 logarithm of the absolute error $\Delta E \vcentcolon = |\langle O \rangle^{(n)}_{\text{VD}} - \bra{\psi}O\ket{\psi}|$ is plotted against the number of copies. BW is only simulated for $N = 4$ and $c \in \{1/8, 1/4, 1/2, 1\}$ due to the demanding computation, and estimated standard errors for its Monte Carlo simulation are represented by shaded areas. By comparing QECR, CR, and BW with the ideal implementation, we observe that errors in the system limits the suppression power of VD, but it is still able to suppress errors for increased number of copies and high $c$. QECR is less effective than CR and BW due to the significant idling errors associated with the repeated preparation of copies, highlighting the importance of preparing $n$ copies in parallel in $n$ nodes.}
	\label{fig: result 1}
\end{figure*}

We decompose all quantum gates involved in the Trotter and VD parts of the circuits into primitives native to trapped-ion platforms (see \Cref{subsec: nat} for details). We model noise in native single (two) qubit gates by applying ideal gates followed by single (two) qubit depolarising error channels. Coherent errors are expected to be negligible in these systems due to precise calibration of control pulses \footnote{Any residual coherent contribution from single-qubit gates can be converted into incoherent errors by the application of twirling techniques.}. Furthermore, we model the error associated with remote Bell state generation via an incoherent mixture of all four Bell states, which is formally equivalent to a single-qubit depolarising channel applied only to one qubit of an ideal Bell pair. We use experimentally realistic values as reference error probabilities, assuming performance typical of leading experiments, specifically $\{p^*_{\text{1Q}},p^*_{\text{2Q}},p^*_{\text{Bell}}\} = \{10^{-4},10^{-3},10^{-2}\}$ for single-qubit gates, two-qubit gates, and remote entanglement generation, respectively. We then conduct a range of simulations considering a range of errors around these nominal values.

Detection errors occur at mid-circuit or final measurements, and correspond to unbiased bit-flips in the measurement outcomes. In our simulations, we fix this to an error probability of $10^{-3}$. Since there are no time constraints in the initialisation of the data qubits in the beginning of the circuit run, we assume this can be performed with negligible error \cite{sotirova2024high}. Moreover, any mid-circuit state preparation that occur during remote entanglement generation lead only to a marginal reduction in the entanglement \emph{rate} with negligible impact on the fidelity. We therefore do not explicitly simulate these state preparation errors. However, the remaining mid-circuit state preparation required to re-prepare new copies in QECR are subject to stricter time constraints, and we model the resulting errors as unbiased bit-flips of probability $10^{-3}$. Assuming that idling errors occur with probability $10^{-5}$ per \SI{1}{\micro\second}, they are simulated by applying single-qubit dephasing channels with error probabilities obtained from the accumulated idling times between consecutive operations.

\subsection{Comparing different implementations \label{subsec: performances}}

We perform direct density matrix simulations to probe expectation values free of shot noise for QECR and CR up to 6 qubits. For BW, we only simulate the case with 4 data qubits and use pure-state Monte Carlo simulations for 4 and 5 copies to circumvent high memory requirements of density matrix simulations. The accuracy of the latter approximate simulation depends on the number of Monte Carlo samples used, as detailed in \Cref{subsec: app}. Therefore, the largest simulation we perform uses $5 \times 4 + 2 = 22$ qubits, which includes 2 ancilla qubits that are initialised in a Bell pair (a bipartite GHZ state). By denoting the error-mitigated expectation value from an ideal or noisy implementation of VD with $n$ copies as $\langle O \rangle^{(n)}_{\text{VD}}$, in \Cref{fig: result 1}, we plot the absolute error from the expectation value obtained from the dominant eigenvector, i.e., $\Delta E \vcentcolon = |\langle O \rangle^{(n)}_{\text{VD}} - \bra{\psi}O\ket{\psi}|$, for an increasing number of copies $n \in \{1,2,3,4,5\}$ in each case. In these simulations, we scale all reference error probabilities by a factor of $c \in 2^{\{ -3, \dots ,2 \}}$, i.e., $\{p_{\text{1Q}}, p_{\text{2Q}}, p_{\text{Bell}}\} = c  \{ p^*_{\text{1Q}},  p^*_{\text{2Q}}, p^*_{\text{Bell}}\}$, but keep idling errors constant. For our BW Monte Carlo simulations, we use shaded areas to show the estimated standard errors in the figure, and we do not consider $c \in \{2,4\}$ due to its excessive computational cost. The regime where quantum error mitigation can be effective is when the circuit error rate is of order 1. We choose the number of Trotter steps consistently such that the circuit error rate of each copy is approximately 1 for the $c=1$ case, and therefore the circuit error rate is approximately $c$ for the different simulations.

\begin{figure*}
    \centering
    \includegraphics[width=0.95\textwidth]{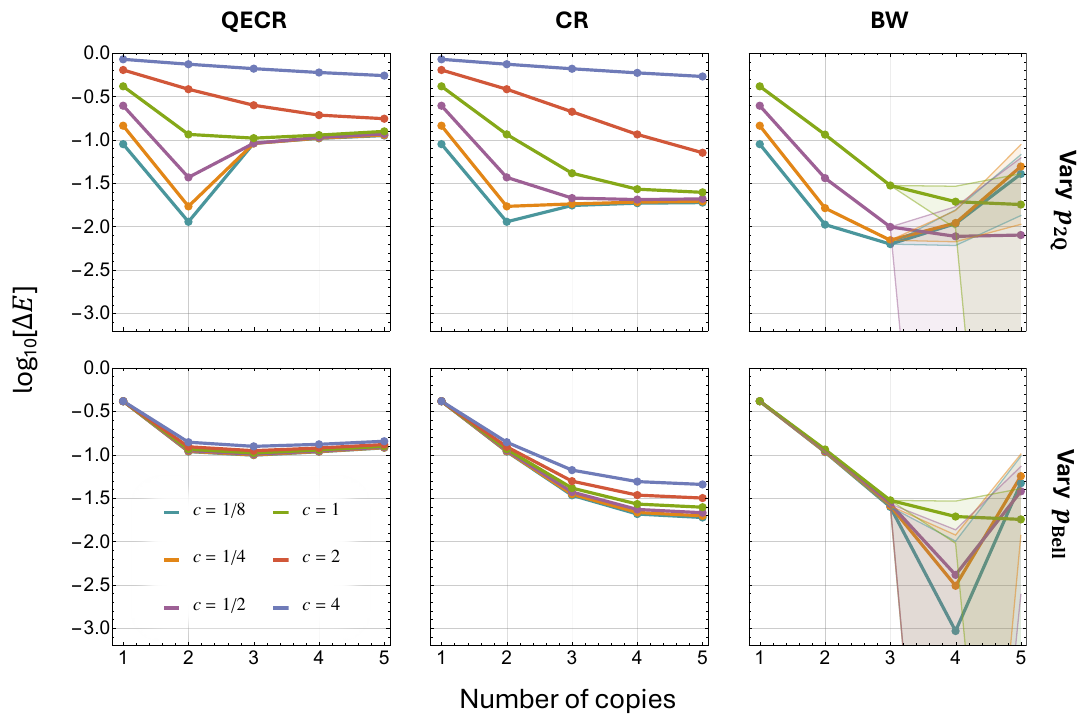}
    \caption{Performance of QECR, CR, and BW (columns) when varying only two-qubit gate error probability as $\{p_{\text{1Q}}, p_{\text{2Q}}, p_{\text{Bell}}\} = \{p^*_{\text{1Q}}, c p^*_{\text{2Q}}, p^*_{\text{Bell}}\}$ (first row) and when varying only entanglement error probability as $\{p_{\text{1Q}}, p_{\text{2Q}}, p_{\text{Bell}}\} = \{p^*_{\text{1Q}}, p^*_{\text{2Q}}, c p^*_{\text{Bell}}\}$ (second row) controlled by the parameter $c \in \{1/8, 1/4, 1/2, 1, 2, 4\}$ (line colors). In each case, the base-10 logarithm of the absolute error $\Delta E \vcentcolon = |\langle O \rangle^{(n)}_{\text{VD}} - \bra{\psi}O\ket{\psi}|$ is plotted against the number of copies. BW is only simulated for $c \in \{1/8, 1/4, 1/2, 1\}$ due to the demanding computation, and estimated standard errors for its Monte Carlo simulation are represented by shaded areas. The robustness of the performance of the three implementations against entanglement errors relative to two-qubit gate errors shows the compatibility of VD with quantum networks.}
    \label{fig: result 2}
\end{figure*}

While theoretical error bounds indeed decrease monotonically as we increase $n$, the absolute error may increase for small $n$ even for an ideal implementation of VD -- but then the absolute error decreases exponentially when the number of copies is sufficiently high~\cite{koczor2021exponential}. This increase for small $n$ can be observed in the case of $N=5$ and $c=4$ of \Cref{fig: result 1} and we further plot in \Cref{subsec: higher} that the absolute error indeed decreases for $n \geq 6$. The variations between $N=4,5,6$ are expected due to the different random instances of the system Hamiltonian considered (see \Cref{subsec: ran}).

The performance of the three practical implementations is affected to various degrees by the presence of errors in the derangement subcircuit. For $N=4$, we observe that both QECR and CR lead to improvements consistently when $n = 2$, but only when $c$ is large enough for $n \geq 3$. In the latter case, their absolute errors plateau, as can be observed for $c \leq 1$ in both implementations. The plateau is around $10^{-1}$ for QECR and around $10^{-2}$ for CR, indicating that the presence of significant idling errors in QECR due to repeated resetting of quantum registers and preparing new copies has a detrimental impact in practice. Similar features can be seen for $N = 5$ and $N = 6$, highlighting the importance of preparing copies in parallel. The performance of BW for $c \leq 1$ is more similar to that of CR than QECR due to the parallel preparation of copies. Furthermore, it improves the absolute error up to $n = 3$ and to lower values compared to CR consistently for all $c \leq 1$. This can be attributed to the reduced circuit depth and two-qubit gate counts given by its constant-depth nature and the replacement of more C-SWAP gates to BSMs in the final layer. Nevertheless, the ancillary GHZ state utilised in BW also introduces additional error sources, and even BW does not show improvements for $n \geq 4$.

\subsection{Analysing robustness against different error sources}
Next, we study how varying the two-qubit gate and entanglement generation error probabilities each affect the performances of the three practical implementations in \Cref{fig: result 2}. Focusing on $N=4$, we set $\{p_{\text{1Q}}, p_{\text{2Q}}, p_{\text{Bell}}\} = \{p^*_{\text{1Q}}, c p^*_{\text{2Q}}, p^*_{\text{Bell}}\}$ in the former and $\{p_{\text{1Q}}, p_{\text{2Q}}, p_{\text{Bell}}\} = \{p^*_{\text{1Q}}, p^*_{\text{2Q}}, c p^*_{\text{Bell}}\}$ in the latter. Since we choose the circuit depth such that the circuit error rate of each copy is close to 1 when $p_{\text{2Q}}=p^*_{\text{2Q}}$ (see \Cref{subsec: ran}), we vary the circuit error rate, along with the errors in the controlled derangement, when varying $p_{\text{2Q}}$.

When varying only two-qubit gate error probabilities, we observe the same qualitative features as the comparison in \Cref{fig: result 1} for $N = 4$. This shows that the relative performance between the three practical implementations depends strongly on two-qubit gate error probabilities. On the other hand, varying only errors in remote entanglement preparation has a minimal impact on the performance of BW when $n \leq 3$, where only at most $2N$ remote Bell pairs are generated, and for all $n$ of QECR and CR. In particular, the absolute error improves monotonically for all $n$ for CR and at least for all $n \leq 4$ for BW in all entanglement error probabilities considered. Thus, while entanglement errors do compromise the performance of VD when applied across a quantum network, the effects are much less concerning than local two-qubit gate errors for realistic error rates, confirming the compatibility between the technique and the hardware.

\section{Conclusion \label{sec: conclusion}}

In this work, we considered practical implementations of VD and compared their performance in realistic scenarios where the qubits and operations involved in VD are subject to errors. We focused on implementation variants of VD that can be applied across a quantum network, including three specific edge cases in space-time tradeoffs. In particular, we considered the standard cyclic rotation (CR), its qubit-efficient variant (QECR), and a depth-efficient variant that can be implemented in a constant circuit depth (in the number of copies) that we term brickwork (BW). We comprehensively detailed implementations of these variants in general and characterised the total numbers of qubits, gate counts and circuit depths required. Furthermore, we explicitly constructed optimised architectures assuming characteristics of networked ion-trap platforms.

We then performed comprehensive numerical simulations assuming realistic noise characteristics of ion trap systems, using density matrix and Monte Carlo simulation methods -- the largest systems we simulated involved $22$ noisy qubits. We confirmed that the accumulation of idling errors in QECR limits its performance and underscores the advantage of CR and BW, which both prepare copies in parallel. While entangling many copies may present a challenge for monolithic quantum devices, the circuits involved are particularly well-suited to networked quantum devices. The primary challenge then becomes the generation of remote entanglement between nodes for the controlled derangement operation, a process we assume will suffer from considerably lower fidelity than can be achieved via local two-qubit gates. However, we observed that, for realistic assumptions, the performance of VD remains primarily limited by the fidelity of local gates, demonstrating its robustness to errors in remote operations. We conclude that, when applied across a quantum network, VD can potentially be effective with more than three copies if the copies are prepared in parallel and reliable implementations of local operations are available.

While our numerical simulations focused on physical gate error characteristics, we highlighted that our implementations of VD can be applied in a networked architecture in an early fault-tolerant setting. In this scenario, each copy stores error-corrected qubits that are manipulated by logical quantum operations~\cite{zimboras2025myths}
and VD is used to mitigate residual logical errors or algorithmic errors due to, e.g., approximate randomised compiling~\cite{campbell2019random,kiumi2024te,kliuchnikov2023shorter,koczor2024sparse}.

Rather than storing each copy in a different node, one can also envision placing individual logical qubits in each node, which is often of interest in distributed logical architectures, where each node can only accommodate a limited number of physical qubits. In this case, the use of one logical ancilla qubit per logical data qubit could be utilised to enable more flexible choices of network connectivities. One can even distribute single logical qubits across physical data qubits placed in distinct nodes. However, these implementations would require logical gates to be applied remotely, leading to higher susceptibility to errors. Combining VD with classical codes applied on quantum architectures with highly biased noise \cite{seis2023improving} could also serve as a resource-efficient means of suppressing errors. Integrating VD into modular early fault-tolerant quantum devices presents a promising direction for future exploration.

\section*{Acknowledgments}
Numerical experiments for this work utilise the Quantum Exact Simulation Toolkit (QuEST) \cite{jones2019quest} via the QuESTlink \cite{jones2020questlink} frontend. Quantum circuits in the main text were illustrated using the Quantikz LaTeX package \cite{kay2023tutorial}. T.A. thanks Tyson Jones for software assistance, such as enabling functionalities in QuESTlink that improved simulation efficiency, and Arjun Rao and Tom Hinde for valuable comments on the manuscript. T.A. acknowledges support by the Oxford-Uehiro Graduate Scholarship Programme. T.A. also acknowledges the use of the University of Oxford Advanced Research Computing (ARC)
facility \cite{richards2015university} in carrying out this work. J.G. thanks the UK Engineering and Physical Sciences Research Council (EPSRC) for funding under the Horizon Europe Guarantee (EP/Y026438/1), and the European Research Council (ERC) for selecting and approving the proposal (ERC Starting Grant: MICRON-QC - 101077098).
B.K. thanks UKRI for the Future Leaders Fellowship Theory to Enable Practical Quantum Advantage (MR/Y015843/1).
The authors also acknowledge funding from the
EPSRC projects Robust and Reliable Quantum Computing (RoaRQ, EP/W032635/1)
and Software Enabling Early Quantum Advantage (SEEQA, EP/Y004655/1). B.K. thanks the University of Oxford for
a Glasstone Research Fellowship and Lady Margaret Hall, Oxford for a Research Fellowship.
This research was funded in part by UKRI (MR/Y015843/1).
For the purpose of Open Access, the author has applied a CC BY public copyright licence
to any Author Accepted Manuscript version arising from this submission.

\appendix

\section{Replacing C-SWAP gates with BSMs \label{sec: rep}}

\begin{figure}
    \centering
    \includegraphics[width=0.5\textwidth]{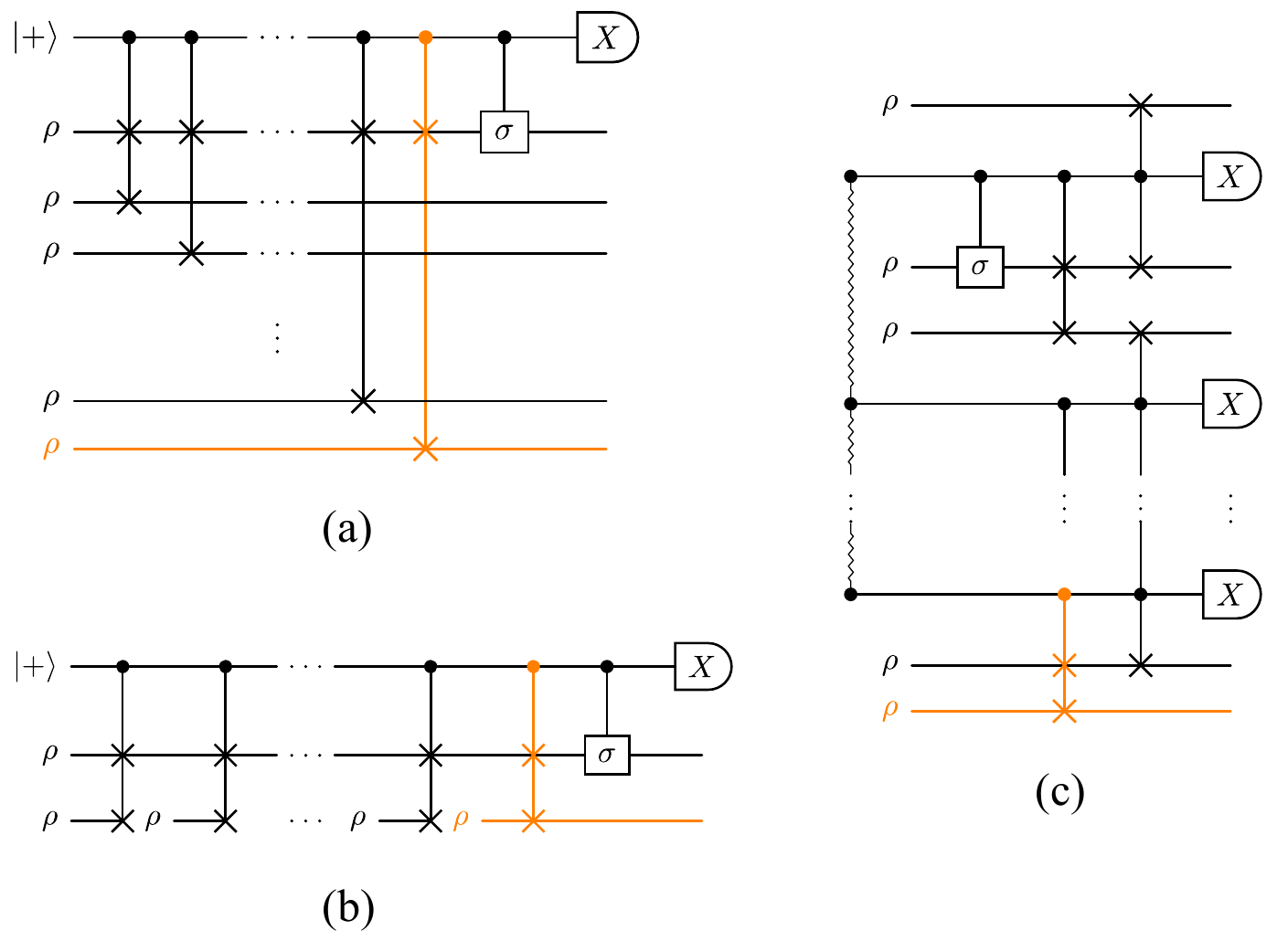}
    \caption{Three implementations of VD before replacing the final layer of C-SWAP gates with BSMs: (a) CR, (b) QECR, and (c) BW. The components highlighted in orange are present (absent) when the number of copies is odd (even). As opposed to the circuits shown in \Cref{fig: vd_circs}, all circuits shown here involve $n-1$ C-SWAP gates and no BSMs. The wavy line across multiple ancilla qubits represents a GHZ state.}
    \label{fig: vd_circs_2}
\end{figure}

For each of the three implementations introduced in \Cref{sec: practical}, all C-SWAP gates on the final layer are replaced by BSMs. The quantum circuits for the three implementations prior to performing this replacement are shown in \Cref{fig: vd_circs_2}. Since the C-SWAP gate is not typically available as a native operation in today's quantum architectures, replacing it with BSMs can lead to significant reductions in the compiled circuit depth. Furthermore, this replacement is most effective for BW, since it involves $\lfloor\frac{n}{2}\rfloor$ transversal C-SWAP gates on the final layer, whereas each of the other methods only has 1. Here, we use the destructive swap test as a minimal example to describe how to arrive at this replacement for VD.

\begin{figure}
    \centering
    \includegraphics[width=0.5\textwidth]{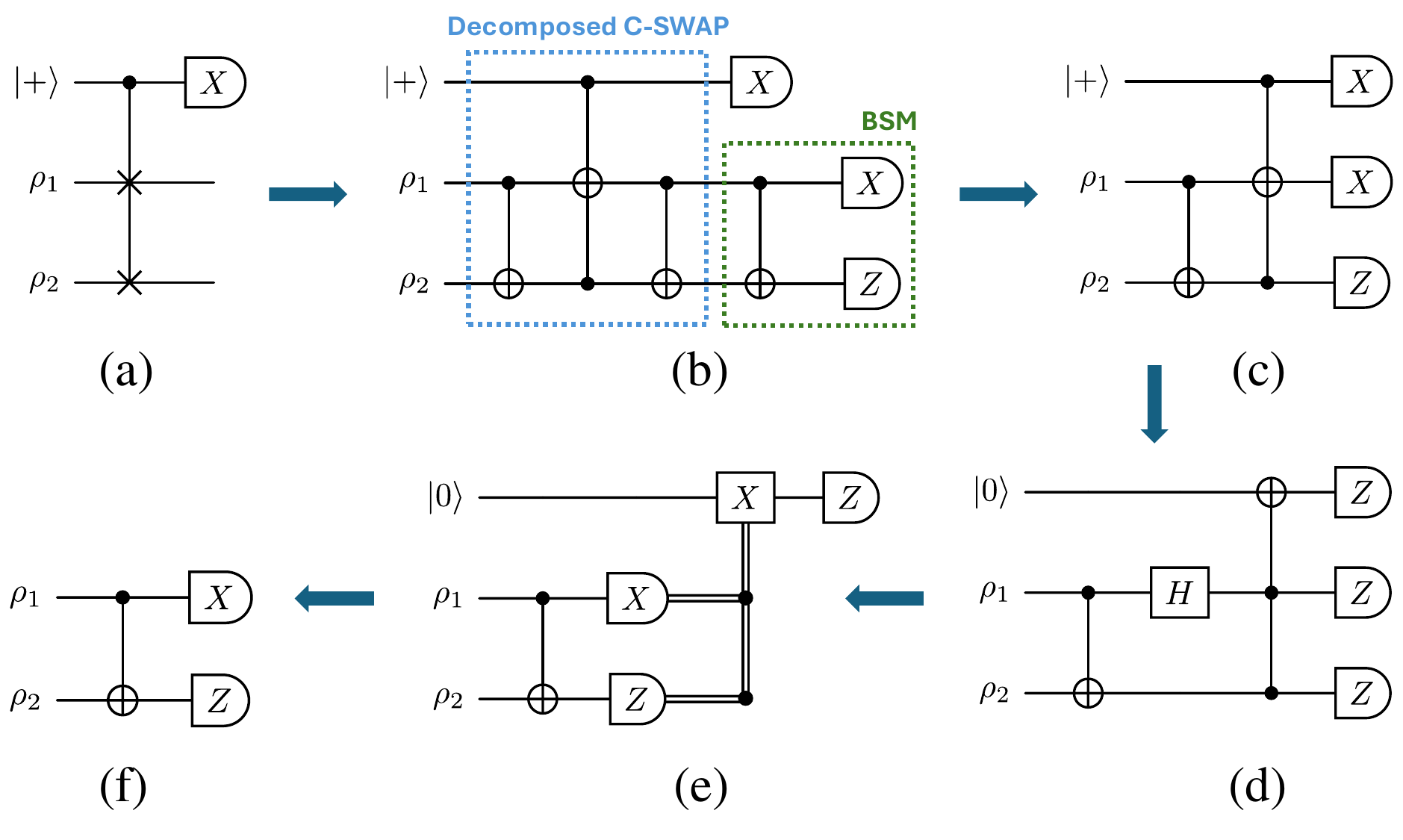}
    \caption{Sketch of equivalence between the non-destructive and destructive swap tests. The same argument allows us to replace the final layer of C-SWAP gates in VD with BSMs.}
    \label{fig: swap test}
\end{figure}

The swap test evaluates the overlap Tr$[\rho_1\rho_2]$ between two quantum states $\rho_1$ and $\rho_2$ using the quantum circuit shown in \Cref{fig: swap test}a. While this is its traditional non-destructive implementation, it is known that the swap test can also be performed destructively as in \Cref{fig: swap test}f, where the ancilla qubit is removed and the C-SWAP gate is replaced by a BSM \cite{garcia2013swap,cincio2018learning}. Their equivalence can be understood by noticing that the state of the ancilla qubit, which is the only qubit that is measured in the end, is unaffected by the application of subsequent operations on all other qubits. In particular, an additional BSM can be applied across the two quantum registers as shown in \Cref{fig: swap test}b. The circuit can be simplified by decomposing the C-SWAP gate into a Toffoli gate sandwiched by a pair of CNOT gates, leading to \Cref{fig: swap test}c following a cancellation of two neighboring CNOT gates. Changing the $X$-basis measurements to $Z$-basis measurements and commuting the Hadamard gates over to the left, we arrive at \Cref{fig: swap test}d, where the ancilla qubit is the target of the Toffoli gate. By the principle of deferred measurement, the Toffoli gate can be changed to an X gate conditioned on classical measurement outcomes as in \Cref{fig: swap test}e. Since, at this stage, only classical information processing is performed on the ancilla qubit, this qubit can be removed from the quantum circuit, noting that the probability of measuring 1 on the ancilla qubit in the original circuit is now translated to whether or not both measurements in \Cref{fig: swap test}f produce 1s in this destructive version.

The same procedure can be applied to each of the C-SWAP gates on the final layer of any VD implementation. Note that bringing the controlled Pauli gates to before the derangement does not change the final measurement outcomes, so we do so to maximise the number of C-SWAP gates that can be replaced by BSMs. When $n$ is even in BW, we also notice that this replacement makes one of the ancilla qubits uninvolved in the derangement, such that we can remove this qubit as highlighted in orange in \Cref{fig: vd_circs}c. This is advantageous in practical implementations, since this implies that VD can be applied with the same number of copies but consuming a smaller GHZ state.

\section{Gate parallelisation with quantum fan-out and fan-in \label{sec: gat}}

The quantum fan-out and fan-in primitives have been utilised to parallelise quantum gates conditioned on ancilla qubits, and they are particularly well suited for modular quantum architectures \cite{haner2021distributed}. Here, we review the fan-out and fan-in primitives to describe how fanning out the ancilla qubit in CR (\Cref{fig: vd_circs_2}a) leads to BW (\Cref{fig: vd_circs_2}c).

Given an arbitrary pure single-qubit state $\ket{\phi^{(1)}}=\alpha\ket{0}+\beta\ket{1}$, a single fan-out expands it to $\ket{\phi^{(2)}}=\alpha\ket{00}+\beta\ket{11}$ for complex amplitudes $\alpha$ and $\beta$. This can be achieved by applying a CNOT gate controlled by the qubit storing $\ket{\phi^{(1)}}$ and targeted on another qubit initialised to $\ket{0}$. The technique can be scaled straightforwardly by applying CNOT gates from the same qubit targeted on additional $\ket{0}$ states, from which we obtain the $s$-qubit state $\ket{\phi^{(s)}}=\alpha\ket{00...0}+\beta\ket{11...1}$ for some $s \geq 2$. When a quantum circuit relies on separable gates that are controlled by a qubit in $\ket{\phi^{(1)}}$, as in the case of VD, fanning it out into $\ket{\phi^{(s)}}$ enables parallel applications of the controlled gates. After using the fanned-out state for parallel gate operations, a reverse fan-in operation brings $\ket{\phi^{(s)}}$ back to $\ket{\phi^{(1)}}$ before its measurement. While the most conceptually simple way to perform fan-in is by applying the same set of CNOT operations as in the fan-out part, the fact that the additional qubits always return to $\ket{0}$ can be utilised to arrive at a destructive fan-in circuit that does not require any two-qubit gate.

In VD and in many other quantum circuits (e.g., the Hadamard test), an ancilla qubit is initially prepared in the $\ket{+}$ state and finally measured in the $X$- (or $Y$-) basis after being used as a control qubit for the main computation. In this case, the ancilla qubit will be fanned out to the Bell state $\ket{\phi^{(2)}}=\frac{1}{\sqrt{2}}(\ket{00}+\ket{11})$ for $s=2$ and more generally to $s$-partite GHZ states. Thus, instead of performing sequential CNOT gates after initializing the ancilla qubit to $\ket{+}$, this fanned out state can be prepared using existing GHZ state preparation techniques that could be more efficient, including those discussed in \Cref{sec: applying}. Similarly, fan-in can also be simplified in this case by noticing that a Z gate conditioned on a classical measurement outcome only induces a bit flip on the final $X$- or $Y$-basis measurement. This means that all the classical communication and the following conditional Z gates are redundant, provided one simply keeps track of the parity between the measurement outcomes. These two simplifications are reflected in the quantum circuit for BW in \Cref{fig: vd_circs}c and \Cref{fig: vd_circs_2}c.

\section{Physical characteristics of a quantum network of trapped ions\label{sec: trapped}}
As a concrete example of a quantum network, we consider for the purposes of this study a trapped-ion quantum architecture in which qubits are encoded in a long-lived transition within some Group-II or Group-II-like ionic species. Such systems exhibit very long coherence times, often via encoding in hyperfine `clock' transitions \cite{langer2005long,olmschenk2007manipulation,sepiol2019,wang2017single} with first-order-insensitivity of frequency to magnetic field. Alternatively, other metastable transitions (e.g. ground-state Zeeman qubits) can be used in combination with ultra-stable magnetic bias fields; in either case decoherence is generally dominated by qubit dephasing. Qubits within each node are assumed to be confined in one or more linear Paul traps such that all-to-all connectivity is achieved either across a single, long chain of ions \cite{wang2020} or via shorter chains of ions, shuttled, split and combined across numerous discrete trapping potentials \cite{kielpinski2002}.

Depending on the qubit type and physical details of the platform, gates may be realised via the application of radiation at optical, microwave or radiofrequency bands. Whatever the mechanism used, this gives rise to a native single qubit gate set consisting of rotations $\text{Rx}(\theta),\text{Ry}(\theta),\text{Rz}(\theta)$, parameterised by angles $\theta$ about the three perpendicular axes $X$, $Y$, and $Z$. While Rx and Ry are noisy physical operations applied over a \SI{1}{\micro\second} timescale \cite{ballance2016high}, Rz is typically an instantaneous noiseless virtual Z gate \cite{mckay2017efficient}. A wide variety of two-qubit gate operations have been demonstrated in recent decades, which achieve entanglement via conditional rotations of a two-qubit state around either equatorial or azimuthal axes \cite{roos2008}. For the purposes of this work, we assume that local entangling operations are achieved via a geometric phase gate which applies a maximally entangling $\text{Rzz}(\theta)$ rotation in \SI{10}{\micro\second} \cite{gaebler2016high,ballance2016high,schafer2018fast}. These single- and two-qubit gates together form a universal gate set within each node.

By enabling remote entanglement, universal quantum computation can be performed across the quantum network. This can be achieved by entangling the internal state of the participating ion in each of two nodes with a state of a photon emitted from that ion. A variety of photonic flying-qubit encodings can be used, including frequency, photon number, polarisation, or time-bin. The photons from the two nodes to be entangled are then directed such that they interfere on a beamsplitter, and their subsequent coincident detection on the far side of this beamsplitter projects the entanglement onto the ions, resulting in an ion-ion Bell state heralded by the coincidence pattern measured. 

Remote entanglement rates are a significant bottleneck in quantum network performance, with leading experiments demonstrating Bell-pair generation rates on the order of \SI{100}{\hertz} \cite{stephenson2020high}. While remote rates are likely to remain lower than those achievable locally, significant increases in performance are anticipated via the use of optical cavities to mediate network ion-photon coupling. In this work we assume that remote entanglement can be achieved at a rate of \SI{10}{\kilo\hertz} with fidelities an order of magnitude lower than local two-qubit gates. Both fidelity and rate assumptions exceed the state-of-the-art in current cavity-mediated remote entanglement experiments \cite{krutyanskiy2023}, but are consistent with realistic engineering assumptions and stand-alone demonstrations of photon collection efficiencies achievable with optical cavities \cite{schupp2021}.

Qubits can be initialised or reset during computation by applying a sequence of cooling techniques, such as Doppler cooling and sideband cooling, followed by optical pumping. The duration of the entire process is on the order of \SI{1}{\micro\second} for the preparation fidelity required for mid-circuit remote entanglement generation ($\sim99\%$); high-fidelity state preparation ($\sim99.99\%$) is typically at least an order-of-magnitude slower, but is only required during initialisation of data qubits $\rho$, before the circuit is applied. The detection of ions is performed by collecting photons from state-dependent fluorescence. The probability of detection errors depends on the duration of the illumination and collection efficiency, but high-fidelity detection is possible in \SI{\sim100}{\micro\second}.

\section{Simulation details}

The results presented in \Cref{sec: numerical} analyses the performance of VD when applied across a quantum network of trapped ions. Here, we describe details of the simulation.

\subsection{Exact simulations \label{subsec: exa}}

In this work, exact simulations are adopted whenever computational resources are manageable for the system sizes of interest. This includes simulations of the ideal VD, and also noisy VD for the single-copy case, CR, QECR, and up to 3 copies of BW. All final measurements in the quantum circuits are computed as exact expectation values of observables instead of sampling random shots.

In an ideal VD, only the copies are noisy and the VD part of the circuit is entirely error-free. Therefore, the simulation can be performed simply by obtaining the density matrix $\rho$ of one noisy copy and computing Tr$[O\rho^n]$ without explicitly constructing the entire circuit. The resource required to perform this computation is not problematic for the system sizes ($N = 4,5,6$) considered in this work.

The single-copy limit of a noisy VD corresponds to not applying VD, so it can be performed in an ancilla-free fashion using $N$ qubits. After preparing a single copy and storing it in memory, a noisy measurement in the $\sigma$-basis is performed for all Pauli term $\sigma$ in the Pauli basis expansion of the observable $O$. The desired quantity Tr$[O\rho]$ can then be obtained by combining the results. which can be done easily for the circuit depths considered.

For QECR, 2 registers of $N$ qubits each store copies onto which VD is applied, with the help of a single ancilla qubit and 2 network qubits. Therefore, the total number of qubits is 15 when $N=6$, which is manageable. Since the simulations for $l$ and $l+1$ copies ($l\geq 2$) only differ by an additional preparation of the copy followed by a layer of C-SWAP gates, we store the intermediate result from the $l^{\text{th}}$ copy so that the simulation of the $l+1^{\text{th}}$ copy does not need to repeat the simulations for all copies up to the $l^{\text{th}}$. Similarly, the circuit for each $l$ is equivalent up to the final layer of controlled Pauli gates for each term in the observable, so we also store intermediate results within each $l$ to minimise the amount of redundant computation. Since most of the gates of the whole circuit are used to prepare each copy, this significantly reduces the time-complexity of the simulation.

An experimental realisation of CR would require a total of $n(N+1)+1$ qubits ($n$ registers with $N+1$ qubits each, including the network qubit, and the ancilla qubit), which is generally much more than the $2N+3$ qubits needed in QECR. However, in our model, CR can be simulated using the same amount of resources as QECR. Since the difference between the two implementations only lie in the errors associated with the preparation of new copies, the simulation of CR can be done by simply switching off these errors that only QECR suffers from.

The simulation of BW gets increasingly challenging with the number of copies for fixed $N$. To alleviate the memory requirement of the simulation, we leave out all network qubits that would otherwise be present in a real experiment. Since we only simulate up to 5 copies, the largest GHZ state that needs to be prepared across the ancilla qubits is a Bell state, which can be prepared directly on the ancilla qubits, so network qubits are not needed for this step. However, errors associated with the other remote operations, namely quantum teleportation and BSMs, must be simulated in slightly modified ways compared to QECR and CR due to the suppression of network qubits. Since we model an entanglement error as a single-qubit depolarising channel acting on one side of an ideal Bell state, this error can be simulated by directly applying the same depolarising channel on the teleported qubit without explicitly performing the teleportation (see Fig.16 of \cite{jnane2022multicore}). Similarly, having a depolarising entanglement error of probability $p_{\text{Bell}}$ in the remote BSM is identical to having a detection error of probability $p_{\text{Bell}}/3$ after the control qubit and $2p_{\text{Bell}}/3$ after the target qubit of the CNOT gate. While idling errors would have accumulated also on the network qubits, those are ignored in our simulations since they are small when network qubits are initialised after the copies are prepared in parallel and are repeatedly reset and re-initialised for remote operations. After removing network qubits, the quantum circuit involves $nN+\lfloor\frac{n-1}{2}\rfloor$ qubits total, such that the largest quantum circuit for BW that we simulate exactly ($N = 4$ and $n = 3$) involves 13 qubits.

\subsection{Approximate simulations \label{subsec: app}}

When exact simulations are not feasible, approximate results can be obtained via Monte Carlo sampling. In this work, we do so for 4 and 5 copies of BW. Instead of applying generally non-unitary operations on density matrices as in exact simulations, in this case we only store state vectors, thus quadratically reducing the memory requirement.

Each non-unitary channel involved in the simulation, excluding measurements, can be defined as a linear combination of tensor products of Pauli operators, so the Monte Carlo simulation can be performed by sampling its terms with probabilities corresponding to its coefficients. For example, a single-qubit depolarising channel of probability $\lambda$ is defined as 
\begin{equation}
    \Delta_\lambda = (1-\lambda) \mathcal{I} + \frac{\lambda}{3}\mathcal{X} + \frac{\lambda}{3}\mathcal{Y} + \frac{\lambda}{3}\mathcal{Z},
\end{equation}
where $\mathcal{I}$, $\mathcal{X}$, $\mathcal{Y}$, and $\mathcal{Z}$ are quantum channels that apply the identity, Pauli $X$, $Y$, and $Z$ operators, respectively. In a Monte Carlo simulation, this channel is replaced by an identity channel with probability $1-\lambda$ and either one of the Pauli X, Y, or Z gates with probability $\lambda/3$.

Given that each of those choices are unitary operations, in a Monte Carlo sampling one needs to only simulate pure states that undergo these random unitary errors. We denote the expected value of the Pauli string $\sigma$ that we obtain from each random circuit variant as the random variable $A_{\rho,n,\sigma}$ whose mean value is the noise expected value as $\mathbb{E}[A_{\rho,n,\sigma}] = \text{Tr}[\sigma\rho^n]$. Ultimately, we want to estimate the ratio that approaches the expectation value with respect to the dominant eigenvector shown in \Cref{eq: vd}, as $\frac{\mathbb{E}[A_{\rho,n,O}]}{\mathbb{E}[A_{\rho,n,\mathbb{I}}]}$, where we defined $A_{\rho,n,O} = \sum_ic_i A_{\rho,n,\sigma_i}$.

The inaccuracy of the simulation, quantified by standard errors, can only be suppressed by taking enough samples. In this work, and in particular for error bars in \Cref{subsec: performances}, the standard error of the biased estimator is estimated by the following procedure: 1) Obtain $M$ samples of $A_{\rho,n,O}$ and $A_{\rho,n,\mathbb{I}}$, where $M$ is chosen as a multiple of 100. 2) Split the samples for $A_{\rho,n,O}$ and $A_{\rho,n,\mathbb{I}}$ into 100 batches each, labeled by $j \in \{1,2,...,100\}$, and compute the mean values $\mu_O^{(j)}$ and $\mu_{\mathbb{I}}^{(j)}$ within each batch. 3) Take the standard deviation between the 100 ratios $\mu_O^{(j)}/\mu_{\mathbb{I}}^{(j)}$.

\begin{figure}[ht]
    \centering
    \subfloat[]{
    \label{subfig: mc_histogram}
        \includegraphics[width=0.48\textwidth]{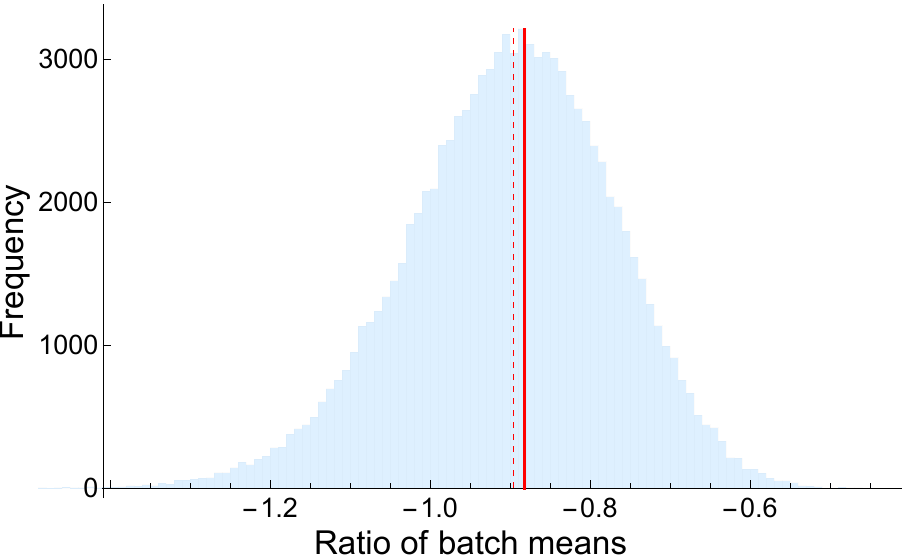}
    }
    \hspace{0.5cm}
    \subfloat[]{
    \label{subfig: mc_error}
        \includegraphics[width=0.48\textwidth]{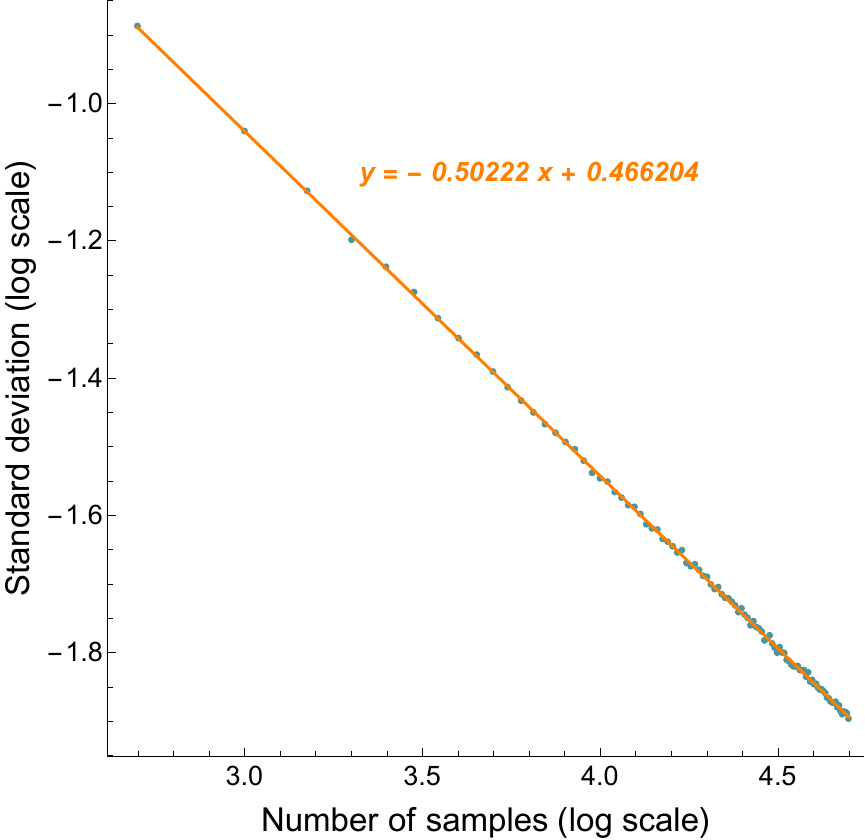}
    }
    \caption{Analysis of data sampled for Monte Carlo simulation with $N = 4$, $n = 2$, and $\{p_{\text{1Q}}, p_{\text{2Q}}, p_{\text{Bell}}\} = \{p^*_{\text{1Q}}, p^*_{\text{2Q}}, p^*_{\text{Bell}}\}$. (a) Distribution of $100000$ ratios of batch means, resampled from a data set of $M = 50000$. The solid and dashed vertical red lines are the exact and approximate simulation results, where the latter case corresponds to the ratio between the average over $M$ samples. (b) Standard deviations (turquoise circles) across $100000$ resampled ratios of batch means, where the number of samples per batch is varied from $500$ to $50000$. A linear fit (orange line) in the log scale reveals an agreement with the standard $1/\sqrt{M}$ scaling from Monte Carlo sampling via the slope and the value of the underlying standard deviation via the y-intercept.}\label{fig: mc}
\end{figure}

To analyse the sampled data, consider the example where $N = 4$, $n = 2$, and $\{p_{\text{1Q}}, p_{\text{2Q}}, p_{\text{Bell}}\} = \{p^*_{\text{1Q}}, p^*_{\text{2Q}}, p^*_{\text{Bell}}\}$. The ratios of batches are distributed as shown in the histogram in \Cref{subfig: mc_histogram} for $M = 50000$. To better visualise the distribution, here we resample from the original data set to obtain $100000$ batches that each contain $M/100 = 500$ random samples, and obtain ratios of batch means. The dashed vertical red line shows the average value obtained from the sampled data, and is situated close to the exact density matrix simulation result, which is the solid vertical line. An agreement with the standard Monte Carlo scaling of $1/\sqrt{M}$ is observed by fitting the standard deviation against $500$ to $50000$ number of samples per $100000$ resampled batches, as shown in \Cref{subfig: mc_error}. The y-intercept of the linear fit shows that the underlying standard deviation of a single sample is $10^{0.466204} \approx 2.92533$, and the number of samples required to obtain some target error can be extrapolated from the fit.

During the Monte Carlo simulation, many sampled quantum circuits correspond to the ideal quantum circuit for small enough error probabilities. Therefore, we save the ideal case in memory and retrieve it whenever the sampled circuit is ideal to avoid performing redundant simulations. Specifically, we store $\ket{\psi_{\text{id}}}^{\otimes n}$, which is the quantum state corresponding to preparing all the ideal copies in parallel. We also store the ideal results $A_{\ket{\psi_{\text{id}}},n,\sigma_i}$ from single runs for all terms indexed by $i$ of the observable. Thus, when the entire circuit is ideal, $A_{\ket{\psi_{\text{id}}},n,\sigma_i}$ is returned without any further computation. If, instead, the copies are ideal but the VD part is noisy, then the prepared ideal copies $\ket{\psi_{\text{id}}}^{\otimes n}$ are retrieved from memory and applied with the noisy VD operations. The entire circuit is simulated otherwise. This is most effective for simulations with low error probabilities, where the probability that the sampled circuit corresponds to the ideal circuit is high.

\subsection{Native-gate decompositions \label{subsec: nat}}

\begin{figure*}
    \centering
    \includegraphics[width=\textwidth]{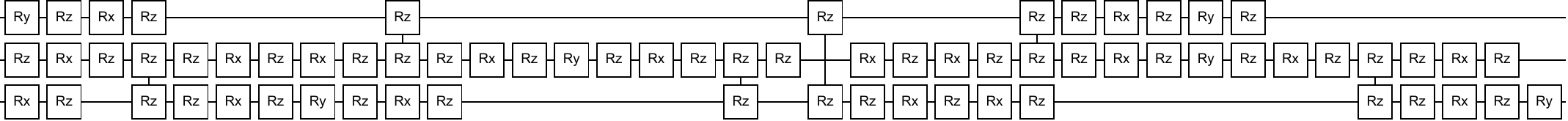}
    \caption{C-SWAP gate decomposed into native gates in the ion trap architecture considered in this work (see \Cref{sec: trapped}), where rotation angles are all integer multiples of $\pi/4$.}
    \label{fig: cswap}
\end{figure*}

For all noisy simulations, we decompose quantum gates into those native in the ion trap hardware described in \Cref{sec: trapped}. Here, we describe how the gates are decomposed, using equal signs to denote equalities up to global phases and order gates temporally such that those on the left are applied before those on the right. Noting that Rz is a virtual Z gate, the Hadamard gate can be applied (on some qubit $q$) with a single physical gate as $\text{H}_q = \text{Rz}_q(\frac{\pi}{2})\text{Rx}_q(\frac{\pi}{2})\text{Rz}_q(\frac{\pi}{2})$. The controlled Pauli gates that correspond to the terms of the observable are each decomposed into single Rzz gates as
\begin{align}
    \begin{split}
        \text{CX}^p_q =& \text{Rz}_p\big(\frac{\pi}{2}\big)\text{Rz}_q\big(\frac{\pi}{2}\big)\text{Rx}_q\big(\frac{\pi}{2}\big)\text{Rz}_q(\pi)\text{Rzz}^p_q\big(\frac{-\pi}{2}\big)\text{H}_q\\
        \text{CY}^p_q =& \text{Rz}_p\big(\frac{\pi}{2}\big)\text{Rx}_q\big(\frac{\pi}{2}\big)\text{Rz}_q\big(\frac{3\pi}{2}\big)\text{Rzz}^p_q\big(\frac{-\pi}{2}\big)\text{Rx}_q\big(\frac{\pi}{2}\big)\text{Rz}_q(\pi)\\
        \text{CZ}^p_q =& \text{Rz}_p\big(\frac{\pi}{2}\big)\text{Rz}_q\big(\frac{\pi}{2}\big)\text{Rzz}^p_q\big(\frac{-\pi}{2}\big),
    \end{split}
\end{align}
where $p$ and $q$ are the control and target qubits of the controlled Pauli gates, respectively. A decomposition of the C-SWAP gate was obtained using a recompiler, as described in detail in \cite{koczor2021exponential}. The decomposition of the C-SWAP gate is shown in \Cref{fig: cswap} for the native gates considered in this work, and includes six Rzz gates.

\begin{figure}
    \centering
    \includegraphics[width=0.48\textwidth]{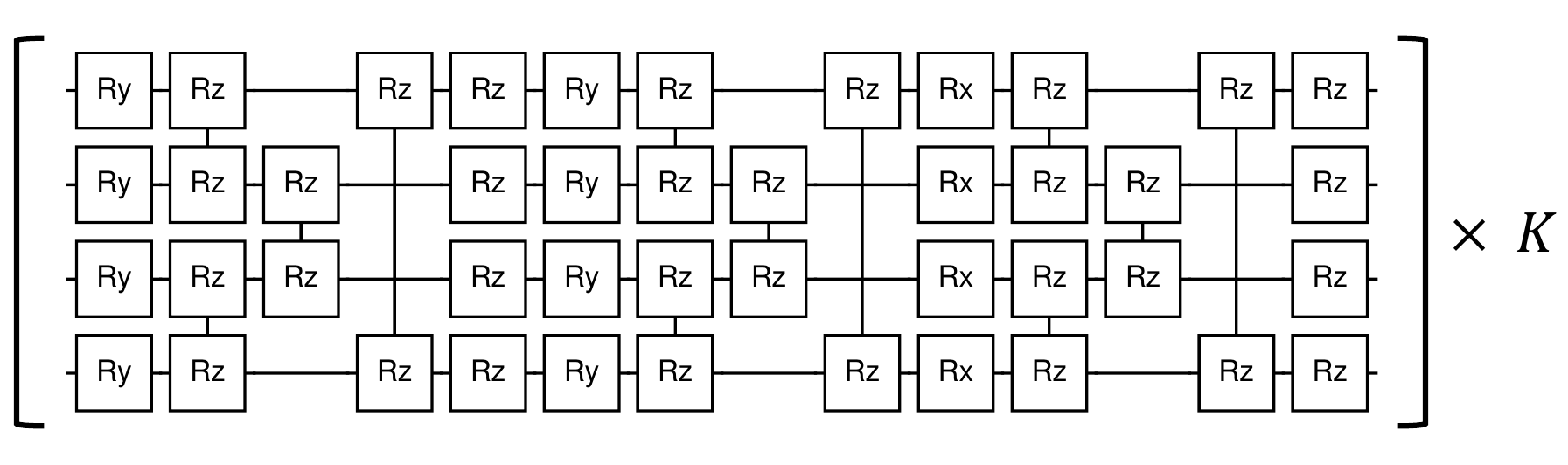}
    \caption{Quantum circuit that prepares each copy of $\rho$ considered in this work when applied onto $\ket{\psi_{\text{init}}}$, shown for $N = 4$. All gates, which are noisy except for the (virtual) Rz gates, are repeated $K$ times, corresponding to $K$ Trotter steps. The preparation of $\ket{\psi_{\text{init}}}$ can be combined with the first layer of Ry gates in the first Trotter steps, since $\ket{\psi_{\text{init}}}$ is a computational basis state.}
    \label{fig: rho}
\end{figure}

\subsection{Artificial gates \label{subsec: art}}

In all numerical simulations that involve explicit simulations of the quantum circuit, the final measurement outcomes are obtained by taking expectation values of observables instead of sampling random shots as would be done in real experiments. For example, considering that we have applied single-qubit gates in the end such that all measurements are in the $Z$-basis, BW requires a measurement of the observable
\begin{equation}
    \bigotimes_i Z_i \bigotimes_j (1-Z_{j_1})(1-Z_{j_2}),
    \label{eq: meas obs 1}
\end{equation}
where $i$ runs over all $\lfloor \frac{n-1}{2} \rfloor$ indices of the ancilla qubits of BW, and $j = (j_1,j_2)$ are $\lfloor \frac{n}{2} \rfloor N$ pairs of indices corresponding to the two qubits involved in individual BSMs. This shows that the number of terms of the observable is $4^{\lfloor \frac{n}{2} \rfloor N}$, which becomes prohibitive for the larger choices of $n$ in this work.

To overcome this issue, we introduce instantaneous and noiseless artificial gates, which do not contribute to the main circuit but are inserted for the sole purpose of converting the measurement observable into a single Pauli string in the numerical simulations. The artificial gates are Toffoli gates (or CCX gates) applied after each conversion to the Bell-basis and any one of the ancilla qubits, such that the outcome of the ancilla qubit is flipped when both outcomes of the BSM are 1. Note that this can be seen as applying the final layer of C-SWAP gates as in \Cref{fig: swap test}d instead of reducing it further down to \Cref{fig: swap test}f. After applying these artificial gates, the full observable in \Cref{eq: meas obs 1} would not be needed, and a single Pauli string $\bigotimes_i Z_i$ would be sufficient.

For QECR and CR, the observable without any artificial gates is given by
\begin{equation}
    Z_\alpha \bigotimes_j (1-Z_{j_{q1}}Z_{j_{a2}})(1-Z_{j_{a1}}Z_{j_{q2}}),
    \label{eq: meas obs 2}
\end{equation}
where $\alpha$ is the index for the ancilla qubit, and $j = (j_{q1},j_{a1},j_{q2},j_{a2})$ correspond to all four qubits involved in the $j^{\text{th}}$ remote BSM according to the notation in \Cref{fig: remote operations}b.

To convert the factor for BSMs in \Cref{eq: meas obs 2} into the same form as that in \Cref{eq: meas obs 1}, we store the parities between qubits $q_1$ and $a_2$ into $q_1$, and between $a_1$ and $q_2$ into $q_2$. This is done by including artificial CNOT gates controlled by $a_2$ ($a_1$) and targeted to $q_1$ ($q_2$) for each of the remote BSMs. 

While the number of terms for simulations of QECR and CR is only $4^N$ due to the fact that only a single set of C-SWAP gates can be replaced by BSMs, applying the artificial Toffoli gates is still crucial. This is due to the complication that network qubits are re-initialised after being used for each remote BSM, while our measurement of \Cref{eq: meas obs 2} requires that all qubits involved be stored in memory after being prepared in the appropriate bases. By inserting the artificial Toffoli gates, along with the artificial CNOT gates, the observable simplifies to $Z_\alpha$, so that a final measurement of the ancilla qubit is sufficient.

\subsection{Random-field Heisenberg model and Trotterisation \label{subsec: ran}}

We consider the example problem to be a one-dimensional spin chain with nearest-neighbor interactions and random magnetic fields, known as the random-field Heisenberg model \cite{luitz2015many,childs2018towards}. Its Hamiltonian is generally given by
\begin{equation}
    \mathcal{H} = J\sum_{j=1}^N \vec{S}_j\cdot\vec{S}_{j+1} + \sum_{j=1}^N h_jZ_j,
    \label{eq: hamiltonian}
\end{equation}
where $\vec{S}_j = (X_j,Y_j,Z_j)$ denotes a vector of Pauli operators acting on site $j$, $J$ is an interaction strength assumed to be uniform across all pairs of interacting spins, and $h_j$ are random magnetic field strengths that are extracted uniformly from the interval $[-h,h]$. We impose periodic boundary conditions, i.e., site $N+1$ is identified with site $1$, and choose $J = h = 1$. Denoting the $N$-dimensional vector of random field strengths as $\vec{h}^{(N)}$, their values in the three system sizes studied in \Cref{subsec: performances} are
\begin{align}
    \vec{h}^{(4)} =& (-0.887, -0.925, -0.72, 0.08)\nonumber\\
    \vec{h}^{(5)} =& (0.206, -0.649, 0.598, -0.826, 0.702)\\
    \vec{h}^{(6)} =& (-0.859, 0.396, -0.354, 0.634, -0.893, 0.198)\nonumber
\end{align}
to three decimal places.

The ideal state $\ket{\psi_{\text{id}}}$ corresponds to a time-evolution under this Hamiltonian, approximated with the Lie-Trotter-Suzuki formula \cite{trotter1959product,suzuki1976generalized}. In this context, an evolution of duration $t$ is separated into $K$ uniform time steps $\delta t = t/K$, within which the time-evolution operator generated by $\mathcal{H}$ can approximately be split into simple quantum rotation gates. Suppose we know how to write the Hamiltonian as $\mathcal{H} = \sum_l \mathcal{H}_l$, where all terms in $\mathcal{H}_l$ in the Pauli basis commute with each other, but $[\mathcal{H}_{l_1}, \mathcal{H}_{l_2}] \neq 0$ for $l_1 \neq l_2$. The ideal state would then be
\begin{equation}
    \ket{\psi_{\text{id}}} = \bigg{(}\prod_l e^{-i\mathcal{H}_l\delta t}\bigg{)}^K \ket{\psi_{\text{init}}} \approx e^{-i\mathcal{H}t} \ket{\psi_{\text{init}}}.
    \label{eq: trotter}
\end{equation}
The initial state $\ket{\psi_{\text{init}}}$ is chosen to be the computational basis state with only site 3 being $\ket{1}$ for all $N$ considered, and we fix $\delta t = 0.01$ such that longer time-evolutions are simulated by increasing $K$. The algorithmic error in \Cref{eq: trotter} depends on $\delta t$, but its choice is unimportant for this work, since we are only interested in suppressing physical errors.

The Hamiltonian in \Cref{eq: hamiltonian} can be grouped into sums of mutually commuting Pauli terms by splitting the sites into an even and odd part. This gives rise to a Trotterised quantum circuit consisting of three layers of two-qubit gates (for three interaction directions) interleaved by single-qubit rotations, as shown in \Cref{fig: rho} for $N = 4$ after compiling for the ion trap quantum hardware described in \Cref{sec: trapped}. Since the initial state is a computational basis state, its preparation is combined with the first layer of Ry gates in the first Trotter step. Within each Trotter step, all single-qubit gates except for the final layer of Rz gates are rotations either by $\frac{\pi}{2}$ or $-\frac{\pi}{2}$. The Rzz gates all share the angle $2J\delta t$, and the final layer of Rz gates have angles given by the random field strengths, $2h_j\delta t$ for $j \in \{1,2,...,N\}$. In line with the applicability of QEM techniques, we choose $K = \lfloor 1/ (3N p^*_{\text{2Q}})\rfloor$ such that the circuit error rate of each copy is approximately 1, where $3N$ is the number of two-qubit gates in a single Trotter step.

\subsection{Performance of VD for $N=5$ and $c = 4$ up to 15 copies} \label{subsec: higher}
\begin{figure}
    \centering
    \includegraphics[width=0.45\textwidth]{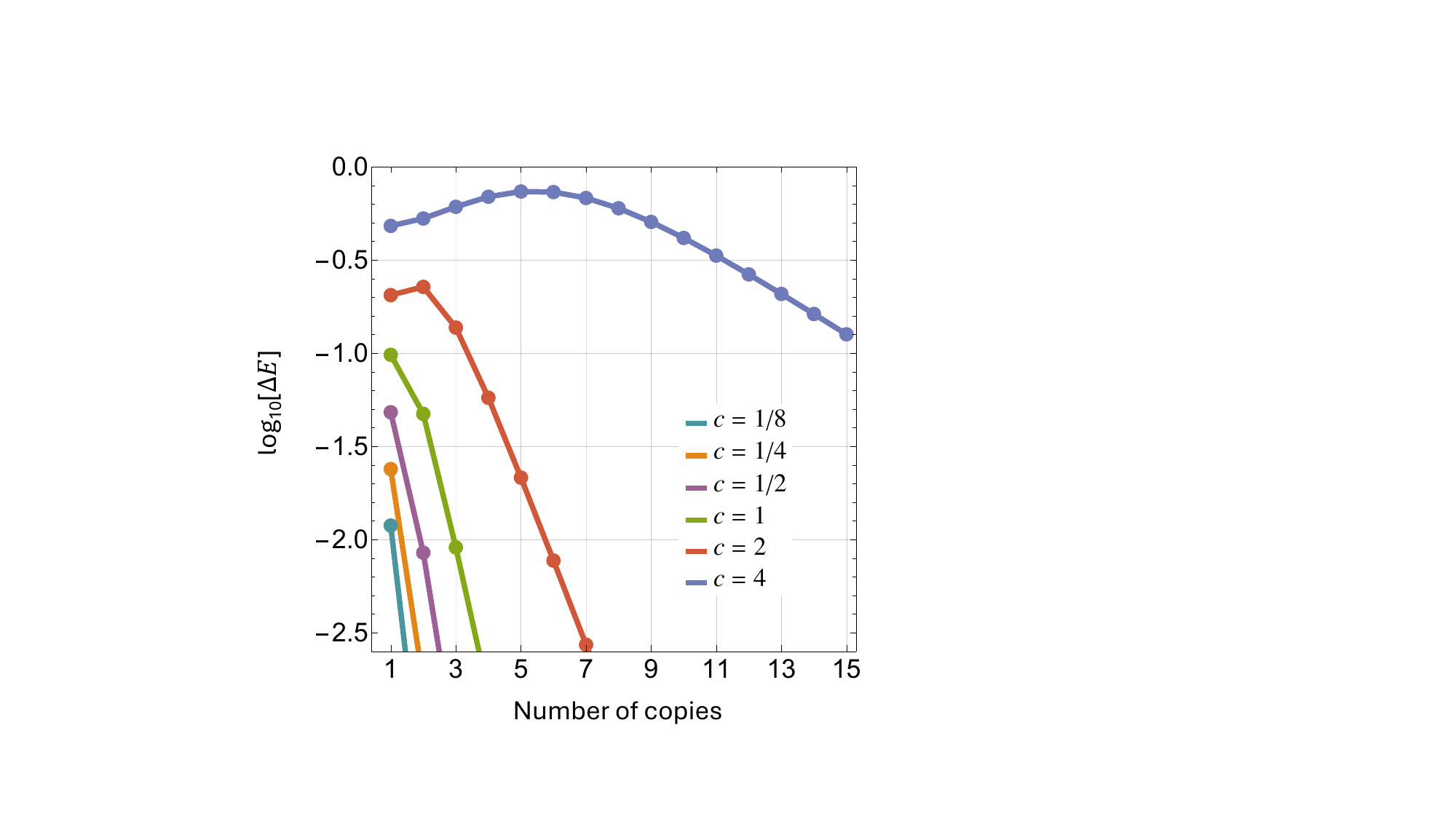}
    \caption{Ideal implementation of VD with $N = 5$ simulated with the same parameter choices as in \Cref{subsec: performances} but up to 15 copies. Even though, as in the corresponding plot in \Cref{fig: result 1}, the absolute error $\Delta E$ in the $c = 4$ case increases in the range $n \in \{1,2,3,4,5\}$, it starts to decrease as $n$ is increased further.}
    \label{fig: higher copies}
\end{figure}

In practice, increasing the number of copies leads to an exponential increase in the sampling cost, such that a value of $n \gtrapprox 10$ may be considered prohibitive. Nevertheless, we may numerically simulate the effect of VD for higher number of copies under an infinite number of circuit runs to study the behaviour of the $N=5$ and $c = 4$ case in \Cref{fig: result 1}, for which the absolute error with the ideal expectation value increases as the number of copies increase in $n \in \{1,2,3,4,5\}$. With the same parameter choices as in \Cref{subsec: performances}, we simulate the ideal implementation of VD up to 15 copies, and the result is shown in \Cref{fig: higher copies}. As we can see, the absolute error begins to decrease from the $6^{\mathrm{th}}$ copy and falls below the $n = 1$ case beyond the $10^{\mathrm{th}}$ copy. This shows that even in this high error rate, the ideal state remains to be the dominant component, but it is less dominant than the rest of the values of $c$ such that the exponential error suppression was not observed from the low number of copies considered in \Cref{fig: result 1}.

\bibliography{references}

\end{document}